\documentclass[a4paper, 11pt]{article}
\pdfoutput=1

\usepackage{jheppub}
\usepackage{afterpage}
\usepackage{multirow}
\usepackage{rotating}

\newcommand{\Dmq}{\Delta m^2}
\newcommand{\dCP}{\delta_\mathrm{CP}}
\newcommand{\Nuc}[2]{{\ensuremath{\mbox{}^{#1}}\text{#2}}}
\newcommand{\eVq}{\ensuremath{\text{eV}^2}}
\newcommand{\diag}{\mathop{\mathrm{diag}}}

\renewcommand{\Im}{\mathop{\mathrm{Im}}}
\newenvironment{pagefigure}{\begin{figure}[!p]}{\afterpage\clearpage\end{figure}}

\title{Global analysis of three--flavour neutrino oscillations:
  synergies and tensions in the determination of $\theta_{23}$,
  $\dCP$, and the mass ordering}

\author[a]{Ivan Esteban,}
\affiliation[a]{Departament de Fis\'{\i}ca Qu\`antica i
  Astrof\'{\i}sica and Institut de Ciencies del Cosmos, Universitat de
  Barcelona, Diagonal 647, E-08028 Barcelona, Spain}
\emailAdd{ivan.esteban@fqa.ub.edu}

\author[a,b,c]{M.~C.~Gonzalez-Garcia,}
\affiliation[b]{Instituci\'o Catalana de Recerca i Estudis
  Avan\c{c}ats (ICREA), Pg. Lluis Companys 23, 08010 Barcelona,
  Spain.}
\affiliation[c]{C.N.~Yang Institute for Theoretical Physics, State
  University of New York at Stony Brook, Stony Brook, NY 11794-3840,
  USA}
\emailAdd{maria.gonzalez-garcia@stonybrook.edu}

\author[d]{Alvaro Hernandez-Cabezudo,}
\affiliation[d]{ Institut f\"ur Kernphysik, Karlsruher Institut f\"ur
  Technologie (KIT), D-76021 Karlsruhe, Germany}
\emailAdd{alvaro.cabezudo@kit.edu}

\author[e]{Michele Maltoni,}
\affiliation[e]{Instituto de F\'{\i}sica Te\'orica UAM/CSIC, Calle de
  Nicol\'as Cabrera 13--15, Universidad Aut\'onoma de Madrid,
  Cantoblanco, E-28049 Madrid, Spain}
\emailAdd{michele.maltoni@csic.es}

\author[d]{Thomas Schwetz}
\emailAdd{schwetz@kit.edu}

\abstract{We present the results of a global analysis of the neutrino
  oscillation data available as of fall 2018 in the framework of three
  massive mixed neutrinos with the goal at determining the ranges of
  allowed values for the six relevant parameters. We describe the
  complementarity and quantify the tensions among the results of the
  different data samples contributing to the determination of each
  parameter. We also show how those vary when combining our global
  likelihood with the $\chi^2$ map provided by Super-Kamiokande for
  their atmospheric neutrino data analysis in the same framework.  The
  best fit of the analysis is for the normal mass ordering with
  inverted ordering being disfavoured with a $\Delta\chi^2 = 4.7\,
  (9.3)$ without (with) SK-atm.  We find a preference for the second
  octant of $\theta_{23}$, disfavouring the first octant with
  $\Delta\chi^2 = 4.4\, (6.0)$ without (with) SK-atm. The best fit for
  the complex phase is $\dCP = 215^\circ$ with CP conservation being
  allowed at $\Delta\chi^2 = 1.5\, (1.8)$. As a byproduct we quantify
  the correlated ranges for the laboratory observables sensitive to
  the absolute neutrino mass scale in beta decay, $m_{\nu_e}$, and
  neutrino-less double beta decay, $m_{ee}$, and the total mass of the
  neutrinos, $\Sigma$, which is most relevant in Cosmology.}

\preprint{IFT-UAM/CSIC-18-112, YITP-SB-18-34}

\begin{document}

\maketitle

\section{Introduction}

Flavour transitions of neutrinos via the energy and distance dependent
neutrino oscillation mechanism~\cite{Pontecorvo:1967fh, Gribov:1968kq}
is a well established phenomenon, which proves that at least two out
of the three neutrinos in the Standard Model must have tiny but
non-zero masses. In this work we revisit the status of three-flavour
neutrino oscillations in view of latest global data.

To fix the convention, the three flavour neutrinos, $\nu_e$, $\nu_\mu$,
$\nu_\tau$, are defined via the weak charged current. They are
expressed as superposition of the three neutrino mass eigen-fields
$\nu_i$ ($i=1,2,3$) with masses $m_i$ via a unitary leptonic mixing
matrix~\cite{Maki:1962mu, Kobayashi:1973fv} by
\begin{equation}
  \nu_\alpha = \sum_{i=1}^3 U_{\alpha i} \nu_i \quad (\alpha = e,\mu,\tau) \,.
\end{equation}
The mixing matrix we parametrize as:
\begin{equation}
  \label{eq:U3m}
  U =
  \begin{pmatrix}
    1 & 0 & 0 \\
    0 & c_{23}  & {s_{23}} \\
    0 & -s_{23} & {c_{23}}
  \end{pmatrix}
  \cdot
  \begin{pmatrix}
    c_{13} & 0 & s_{13} e^{-i\dCP} \\
    0 & 1 & 0 \\
    -s_{13} e^{i\dCP} & 0 & c_{13}
  \end{pmatrix}
  \cdot
  \begin{pmatrix}
    c_{21} & s_{12} & 0 \\
    -s_{12} & c_{12} & 0 \\
    0 & 0 & 1
  \end{pmatrix}
   \cdot \mathcal{P}
\end{equation}
where $c_{ij} \equiv \cos\theta_{ij}$ and $s_{ij} \equiv
\sin\theta_{ij}$.  The angles $\theta_{ij}$ can be taken without loss
of generality to lie in the first quadrant, $\theta_{ij} \in [0,
  \pi/2]$, and the phase $\dCP \in [0, 2\pi]$. Values of $\dCP$
different from 0 and $\pi$ imply CP violation in neutrino oscillations
in vacuum~\cite{Cabibbo:1977nk, Bilenky:1980cx, Barger:1980jm}.
$\mathcal{P}$ is a diagonal matrix which is the identity if neutrinos
are Dirac fermions and it contains two additional phases,
$\mathcal{P}=\diag(e^{i\alpha_1}, e^{i\alpha_2},1)$, if they are
Majorana fermions. The Majorana phases $\alpha_1$ and $\alpha_2$ play
no role in neutrino oscillations~\cite{Bilenky:1980cx,
  Langacker:1986jv}.

In this convention there are two non-equivalent orderings for the
neutrino masses, namely normal ordering (NO) with $m_1 < m_2 < m_3$,
and inverted ordering (IO) with $m_3 < m_1 < m_2$. Furthermore the
data show a hierarchy between the mass splittings, $\Dmq_{21} \ll
|\Dmq_{31}| \simeq |\Dmq_{32}|$ with $\Dmq_{ij} \equiv m_i^2 - m_j^2$.
In this work we follow the convention from
Ref.~\cite{Gonzalez-Garcia:2014bfa} and present our results for both,
NO and IO, using the smallest and largest mass splittings. The
smallest one is always $\Dmq_{21}$, while the largest one we denote by
$\Dmq_{3\ell}$, with $\ell=1$ for NO and $\ell=2$ for IO.  Hence,
\begin{equation}
  \Dmq_{3\ell} = \left\{
  \begin{array}{l}
    \Dmq_{31} > 0 \quad\text{for NO}\,,\\
    \Dmq_{32} < 0 \quad\text{for IO}\,.
  \end{array} \right.
\end{equation}

Due to the wealth of experiments exploring neutrino oscillations, we
are in the situation that a given parameter is determined by several
measurements. Therefore, combined analyses such as the one presented
below are an important tool to extract the full information on
neutrino oscillation parameters. This is especially true for open
questions, such as the octant of $\theta_{23}$, the type of the
neutrino mass ordering, and the status of the complex phase $\dCP$,
where some hints are emerging due to significant synergies between
different experiments. However, also for parameters describing
dominant oscillations, a significantly more accurate determination
emerges by the combination of complementary data sets, such as for
example for $|\Dmq_{3\ell}|$.

We present below the global fit NuFIT-4.0, updating our previous
analyses~\cite{GonzalezGarcia:2012sz, Gonzalez-Garcia:2014bfa,
  Esteban:2016qun}.  $\Delta\chi^2$ maps and future updates of this
analysis will be made available at the NuFIT website~\cite{nufit}. For
other recent global fits see~\cite{deSalas:2017kay, Capozzi:2018ubv}.

\section{Global analysis: determination of oscillation parameters}
\label{sec:global}

\subsection{Data samples analyzed}
\label{sec:data}

The analysis presented below uses data available up to fall 2018.  A
complete list of the used data including references can be found in
appendix~\ref{sec:appendix-data}. Here we give a brief overview of
recent updates and mention changes with respect to our previous
analysis~\cite{Esteban:2016qun}.

We include latest data from the MINOS~\cite{Adamson:2013whj,
  Adamson:2013ue}, T2K~\cite{Abe:2017vif, Abe:2018wpn}, and
NOvA~\cite{Adamson:2017gxd, NOvA:2018gge} long-baseline accelerator
experiments from $\nu_\mu$ disappearance and $\nu_\mu\to\nu_e$
appearance channels, both for neutrino and anti-neutrino beam
modes. In particular, T2K and NOvA have presented updated results at
the Neutrino18 conference, including also first data on anti-neutrinos
from NOvA, whose impact will be discussed below.

Concerning reactor neutrino experiments, the fit of data with
baselines in the km range (medium baseline, MBL) is completely
dominated by modern experiments, most importantly by Daya
Bay~\cite{Adey:2018zwh}, with subleading contributions from
RENO~\cite{Bak:2018ydk} and Double Chooz~\cite{Abe:2014bwa}. Moreover,
those experiments are entirely based on relative spectra from
detectors at different baselines, and are therefore largely
independent of reactor neutrino flux predictions. In view of the
unclear situation of reactor flux predictions and reactor data at very
short baselines (see, \textit{e.g.}, Ref.~\cite{Dentler:2017tkw} for a
recent discussion), we decided to include only the modern MBL reactor
experiments Daya Bay, RENO, and Double Chooz. For the analysis of
KamLAND long-baseline reactor data we replaced predicted neutrino
fluxes by the spectrum measured in Daya Bay near
detectors~\cite{An:2016srz}, which makes also our KamLAND analysis
largely independent of flux predictions.

Our solar neutrino data includes previous data from radio-chemical and
the SNO experiments, as well as updated exposures from
Super-Kamiokande and Borexino, see appendix~\ref{sec:appendix-data}
for the detailed list and references.\footnote{We do not include here
  the latest data release from Borexino~\cite{borexino:2018}, which is
  expected to have a very small impact on the determination of
  oscillation parameters. These data will be included in future
  updates of our global fit.}

Atmospheric neutrino data generically are difficult to analyze outside
the experimental collaborations. We present below two separate global
analyses, depending on the used atmospheric neutrino data. Our default
analysis makes use of IceCube/DeepCore 3-year
data~\cite{Aartsen:2014yll} which can be re-analyzed using the
information provided by the collaboration~\cite{deepcore:2016}.
Especially in the context of the mass ordering determination,
atmospheric neutrino data from Super-Kamiokande 1-4~\cite{Abe:2017aap}
seems to provide important information.  Unfortunately there is not
enough information available to reproduce these results outside the
collaboration. However, Super-Kamiokande has published the results of
their analysis in the form of a tabulated $\chi^2$
map~\cite{SKatm:data2018}, which we can combine with our global
analysis. We will show the results of this combination as an
alternative global fit.  A detailed discussion of atmospheric neutrino
data, including also the potential impact of an alternative IceCube
analysis~\cite{Aartsen:2017nmd} will be presented in
section~\ref{sec:atm-data}.

\subsection{Summary of global fit results}
\label{sec:oscparam}

The results of our global fit are displayed in
fig.~\ref{fig:chisq-glob} (one-dimensional $\Delta\chi^2$ curves) and
fig.~\ref{fig:region-glob} (two-dimensional projections of confidence
regions).  In table~\ref{tab:bfranges} we give the best fit values as
well as $1\sigma$ and $3\sigma$ confidence intervals for the
oscillation parameters. We show two versions of the results. The
default analysis is without Super-Kamiokande atmospheric neutrino data
(SK-atm), and contains all the data for which a fit can be
performed. For the alternative analysis, we add the pre-calculated
$\Delta\chi^2$ table from SK-atm provided by the collaboration to our
global fit, in order to illustrate the potential impact of these
data. Let us summarize here the main features of the global fit
result. More detailed discussions about how certain features emerge
will be given in the following sections.

\begin{pagefigure}\centering
  \includegraphics[width=0.86\textwidth]{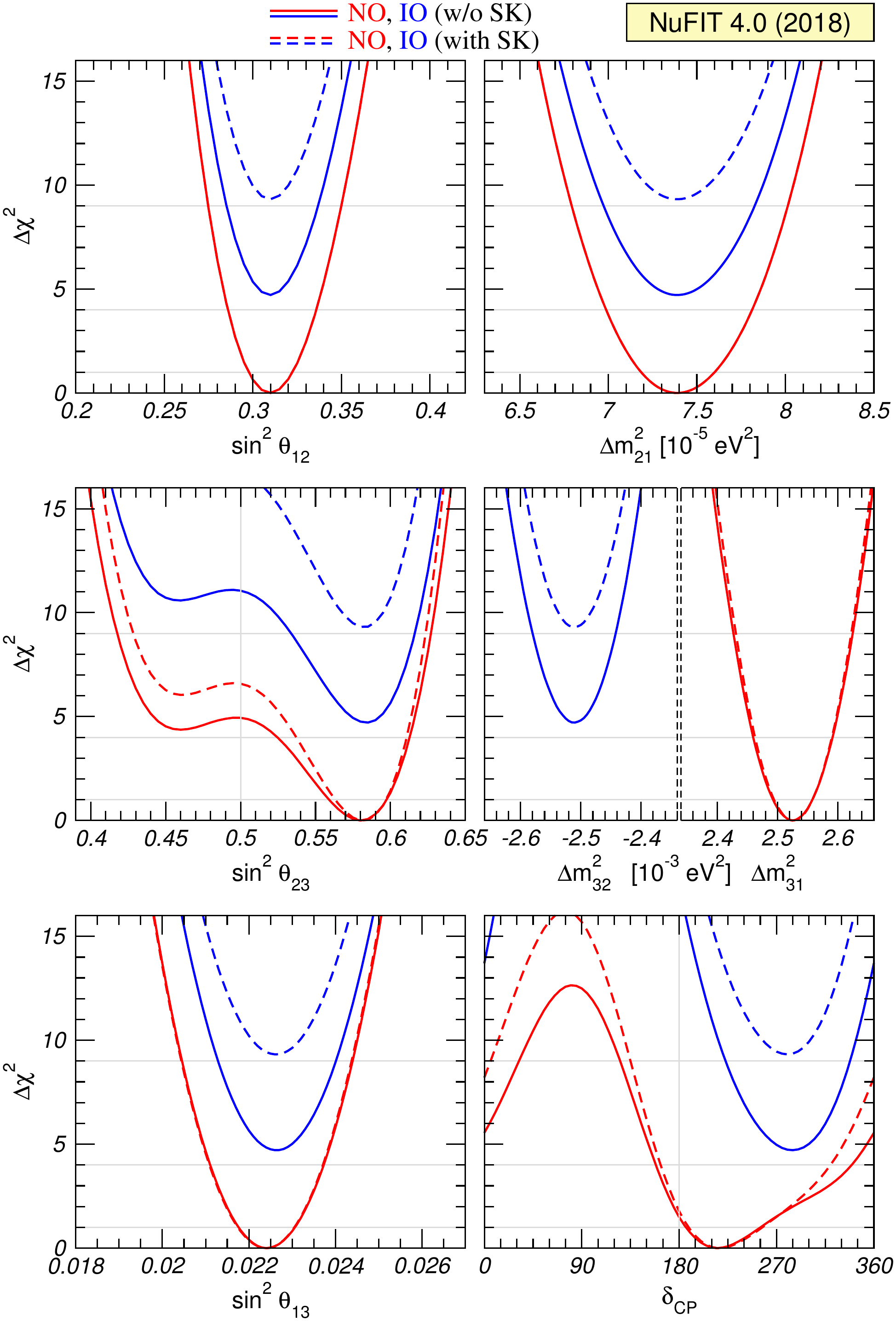}
  \caption{Global $3\nu$ oscillation analysis. We show $\Delta\chi^2$
    profiles minimized with respect to all undisplayed parameters. The
    red (blue) curves correspond to Normal (Inverted) Ordering. Solid
    (dashed) curves are without (with) adding the tabulated SK-atm
    $\Delta\chi^2$.  Note that as atmospheric mass-squared splitting
    we use $\Dmq_{31}$ for NO and $\Dmq_{32}$ for IO.}
  \label{fig:chisq-glob}
\end{pagefigure}

\begin{pagefigure}\centering
  \includegraphics[width=0.81\textwidth]{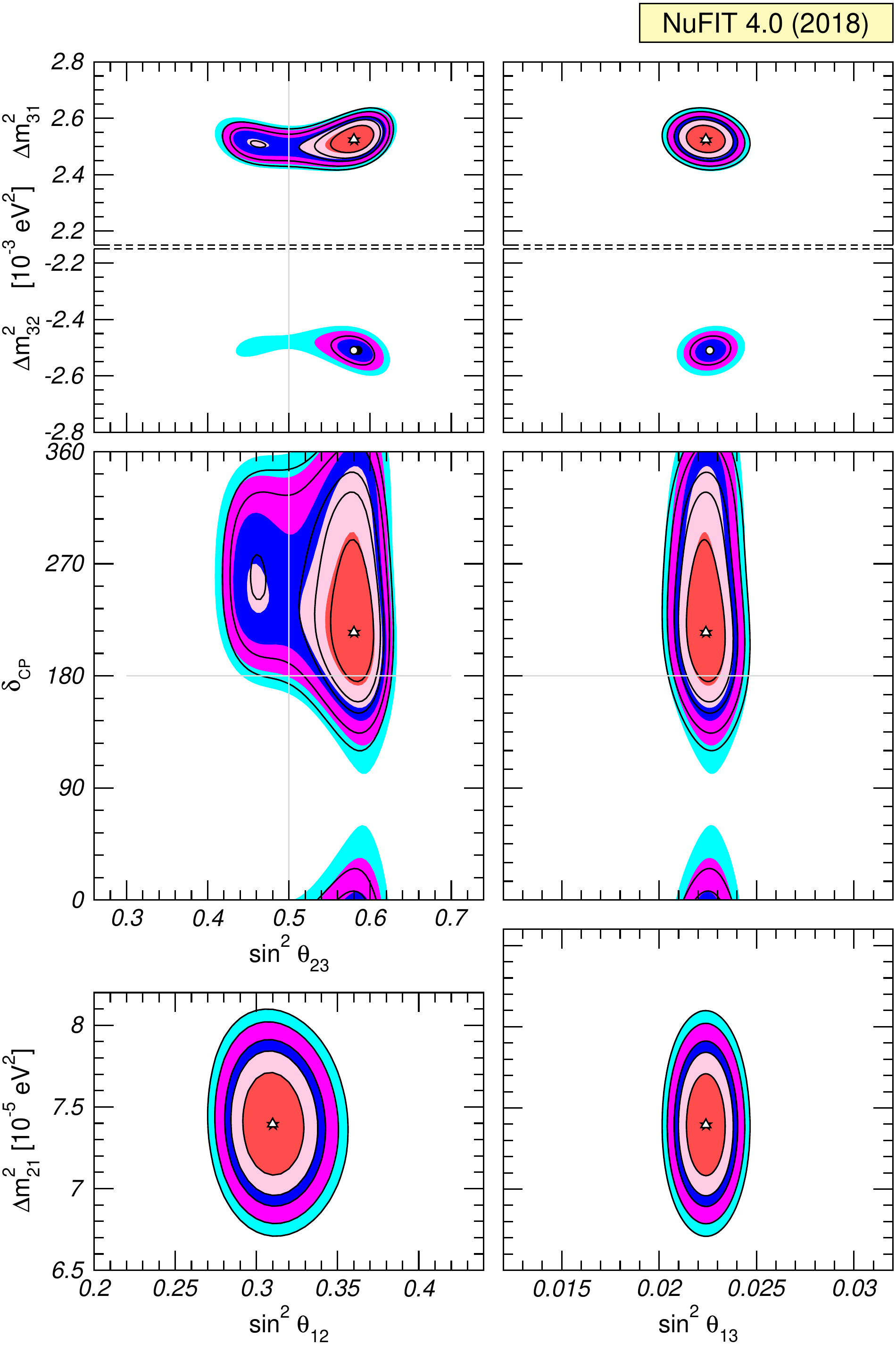}
  \caption{Global $3\nu$ oscillation analysis. Each panel shows the
    two-dimensional projection of the allowed six-dimensional region
    after minimization with respect to the undisplayed parameters. The
    regions in the four lower panels are obtained from $\Delta\chi^2$
    minimized with respect to the mass ordering. The different
    contours correspond to $1\sigma$, 90\%, $2\sigma$, 99\%, $3\sigma$
    CL (2 dof). Coloured regions (black contour curves) are without
    (with) adding the tabulated SK-atm $\Delta\chi^2$. Note that as
    atmospheric mass-squared splitting we use $\Dmq_{31}$ for NO and
    $\Dmq_{32}$ for IO.}
  \label{fig:region-glob}
\end{pagefigure}

\begin{table}\centering
  \begin{footnotesize}
    \begin{tabular}{l|l|cc|cc}
      \hline\hline
      &
      & \multicolumn{2}{c|}{Normal Ordering (best fit)}
      & \multicolumn{2}{c}{Inverted Ordering ($\Delta\chi^2=4.7$)}
      \\
      \hline
      && bfp $\pm 1\sigma$ & $3\sigma$ range
      & bfp $\pm 1\sigma$ & $3\sigma$ range
      \\
      \hline
      \rule{0pt}{4mm}\ignorespaces
      &
      $\sin^2\theta_{12}$
      & $0.310_{-0.012}^{+0.013}$ & $0.275 \to 0.350$
      & $0.310_{-0.012}^{+0.013}$ & $0.275 \to 0.350$
      \\[1mm]
      &
      $\theta_{12}/^\circ$
      & $33.82_{-0.76}^{+0.78}$ & $31.61 \to 36.27$
      & $33.82_{-0.76}^{+0.78}$ & $31.61 \to 36.27$
      \\[3mm]
      &
      $\sin^2\theta_{23}$
      & $0.580_{-0.021}^{+0.017}$ & $0.418 \to 0.627$
      & $0.584_{-0.020}^{+0.016}$ & $0.423 \to 0.629$
      \\[1mm]
      &
      $\theta_{23}/^\circ$
      & $49.6_{-1.2}^{+1.0}$ & $40.3 \to 52.4$
      & $49.8_{-1.1}^{+1.0}$ & $40.6 \to 52.5$
      \\[3mm]
      &
      $\sin^2\theta_{13}$
      & $0.02241_{-0.00065}^{+0.00065}$ & $0.02045 \to 0.02439$
      & $0.02264_{-0.00066}^{+0.00066}$ & $0.02068 \to 0.02463$
      \\[1mm]
      \multirow{10}{*}{
        \begin{rotate}{90} without SK-atm \hspace*{1cm}\end{rotate}}
      &
      $\theta_{13}/^\circ$
      & $8.61_{-0.13}^{+0.13}$ & $8.22 \to 8.99$
      & $8.65_{-0.13}^{+0.13}$ & $8.27 \to 9.03$
      \\[3mm]
      &
      $\dCP/^\circ$
      & $215_{-29}^{+40}$ & $125 \to 392$
      & $284_{-29}^{+27}$ & $196 \to 360$
      \\[3mm]
      &
      $\dfrac{\Dmq_{21}}{10^{-5}~\eVq}$
      & $7.39_{-0.20}^{+0.21}$ & $6.79 \to 8.01$
      & $7.39_{-0.20}^{+0.21}$ & $6.79 \to 8.01$
      \\[3mm]
      &
      $\dfrac{\Dmq_{3\ell}}{10^{-3}~\eVq}$
      & $+2.525_{-0.032}^{+0.033}$ & $+2.427 \to +2.625$
      & $-2.512_{-0.032}^{+0.034}$ & $-2.611 \to -2.412$
      \\[3mm]
      \hline\hline
      &
      & \multicolumn{2}{c|}{Normal Ordering (best fit)}
      & \multicolumn{2}{c}{Inverted Ordering ($\Delta\chi^2=9.3$)}
      \\
      \hline
      && bfp $\pm 1\sigma$ & $3\sigma$ range
      & bfp $\pm 1\sigma$ & $3\sigma$ range
      \\     \hline
      \rule{0pt}{4mm}\ignorespaces
      &
      $\sin^2\theta_{12}$
      & $0.310_{-0.012}^{+0.013}$ & $0.275 \to 0.350$
      & $0.310_{-0.012}^{+0.013}$ & $0.275 \to 0.350$
      \\[1mm]
      &
      $\theta_{12}/^\circ$
      & $33.82_{-0.76}^{+0.78}$ & $31.61 \to 36.27$
      & $33.82_{-0.75}^{+0.78}$ & $31.62 \to 36.27$
      \\[3mm]
      &
      $\sin^2\theta_{23}$
      & $0.582_{-0.019}^{+0.015}$ & $0.428 \to 0.624$
      & $0.582_{-0.018}^{+0.015}$ & $0.433 \to 0.623$
      \\[1mm]
      &
      $\theta_{23}/^\circ$
      & $49.7_{-1.1}^{+0.9}$ & $40.9 \to 52.2$
      & $49.7_{-1.0}^{+0.9}$ & $41.2 \to 52.1$
      \\[3mm]
      \multirow{10}{*}{\begin{rotate}{90} with SK-atm \hspace*{1cm}\end{rotate}}
      &
      $\sin^2\theta_{13}$
      & $0.02240_{-0.00066}^{+0.00065}$ & $0.02044 \to 0.02437$
      & $0.02263_{-0.00066}^{+0.00065}$ & $0.02067 \to 0.02461$
      \\[1mm]
      &
      $\theta_{13}/^\circ$
      & $8.61_{-0.13}^{+0.12}$ & $8.22 \to 8.98$
      & $8.65_{-0.13}^{+0.12}$ & $8.27 \to 9.03$
      \\[3mm]
      &
      $\dCP/^\circ$
      & $217_{-28}^{+40}$ & $135 \to 366$
      & $280_{-28}^{+25}$ & $196 \to 351$
      \\[3mm]
      &
      $\dfrac{\Dmq_{21}}{10^{-5}~\eVq}$
      & $7.39_{-0.20}^{+0.21}$ & $6.79 \to 8.01$
      & $7.39_{-0.20}^{+0.21}$ & $6.79 \to 8.01$
      \\[3mm]
      &
      $\dfrac{\Dmq_{3\ell}}{10^{-3}~\eVq}$
      & $+2.525_{-0.031}^{+0.033}$ & $+2.431 \to +2.622$
      & $-2.512_{-0.031}^{+0.034}$ & $-2.606 \to -2.413$
      \\[2mm]
      \hline\hline
    \end{tabular}
  \end{footnotesize}
  \caption{Three-flavour oscillation parameters from our fit to global
    data. The numbers in the 1st (2nd) column are obtained assuming NO
    (IO), \textit{i.e.}, relative to the respective local minimum.
    Note that $\Dmq_{3\ell} \equiv \Dmq_{31} > 0$ for NO and
    $\Dmq_{3\ell} \equiv \Dmq_{32} < 0$ for IO. The results shown in
    the upper (lower) table are without (with) adding the tabulated
    SK-atm $\Delta\chi^2$.}
  \label{tab:bfranges}
\end{table}

Except for $\sin^2\theta_{23}$ and $\dCP$ the $\Delta\chi^2$ shapes
are close to parabolic, indicating that the $\chi^2$ approximation for
the distribution should hold to good accuracy. The Monte Carlo studies
performed in Refs.~\cite{Esteban:2016qun, Elevant:2015ska} indicate
that also for $\sin^2\theta_{23}$, $\dCP$ and the mass ordering the
$\chi^2$ approximation gives a reasonable estimate of the
corresponding confidence level. Therefore, the $\Delta\chi^2$ values
given below can be converted into an approximate number of standard
deviations by the $\sqrt{\Delta\chi^2}$ rule.

Defining the $3\sigma$ relative precision of the parameter by
$2(x^\text{up} - x^\text{low}) / (x^\text{up} + x^\text{low})$, where
$x^\text{up}$ ($x^\text{low}$) is the upper (lower) bound on a
parameter $x$ at the $3\sigma$ level, we obtain the following
$3\sigma$ relative precisions (marginalizing over ordering):
\begin{equation}
  \label{eq:precision}
  \begin{array}{l@{\,,\qquad}l@{\,,\qquad}l}
    14\% \, (\theta_{12})\, &   8.9\% \, (\theta_{13})\,
    &  27\,[24]\% \, (\theta_{23}) \,, \\
    16\% \,(\Dmq_{21}) \, & 7.8\, [7.6]\% \,(|\Dmq_{3\ell}|)\,
    & 100\,[92]\%\,(\dCP) \,,\\
  \end{array}
\end{equation}
where the numbers between brackets show the impact of including SK-atm
in the precision of that parameter determination. We notice that as
$\Delta\chi^2$ shape for $\dCP$ is clearly not gaussian this
evaluation of its ``precision'' can only be taken as indicative.

Altogether the status of mass ordering discrimination, determination
of $\sin^2\theta_{23}$, and the leptonic CP phase $\dCP$ can be
summarized as follows:
\begin{itemize}
\item The best fit is for the normal mass ordering. Inverted ordering
  is disfavoured with a $\Delta\chi^2 = 4.7 \, (9.3)$ without (with)
  SKatm.

\item We obtain preference for the second octant of $\theta_{23}$,
  with the best fit point located at $\sin^2\theta_{23} =
  0.58$. Values with $\sin^2\theta_{23} \le 0.5$ are disfavoured with
  $\Delta\chi^2 = 4.4 \, (6.0)$ without (with) SK-atm.

\item The best fit for the complex phase is at $\dCP = 215^\circ$.
  Compared to previous results (\textit{e.g.}, NuFIT
  3.2~\cite{nufit}), the allowed range is pushed towards the CP
  conserving value of $180^\circ$, which now is only disfavoured with
  $\Delta\chi^2 = 1.5$ (1.8) without (with) SK-atm.
\end{itemize}

In table~\ref{tab:bfranges} we give the best fit values and confidence
intervals for both mass orderings, relative to the local best fit
points in each ordering. The global confidence intervals
(marginalizing also over the ordering) are identical to the ones for
normal ordering, which have also been used in
eq.~\eqref{eq:precision}.  The only exception to this statement is
$\Dmq_{3\ell}$ in the analysis without SK-atm; in this case a
disconnected interval would appear above $2\sigma$ corresponding to
negative values of $\Dmq_{3\ell}$ (\textit{i.e.}, inverted
ordering). Altogether we derive the following $3\sigma$ ranges on the
magnitude of the elements of the leptonic mixing matrix:
\begin{align}
  |U|_{3\sigma}^\text{\,w/o\,SK\,atm} &=
  \begin{pmatrix}
    0.797 \to 0.842 &\qquad
    0.518 \to 0.585 &\qquad
    0.143 \to 0.156
    \\
    0.233 \to 0.495 &\qquad
    0.448 \to 0.679 &\qquad
    0.639 \to 0.783
    \\
    0.287 \to 0.532 &\qquad
    0.486 \to 0.706 &\qquad
    0.604 \to 0.754
  \end{pmatrix}   \label{eq:umatrix}
  \\[1mm] \nonumber
  |U|_{3\sigma}^\text{\, w\,SK\,atm} &=
  \begin{pmatrix}
    0.797 \to 0.842 &\qquad
    0.518 \to 0.585 &\qquad
    0.143 \to 0.156
    \\
    0.235 \to 0.484 &\qquad
    0.458 \to 0.671 &\qquad
    0.647 \to 0.781
    \\
    0.304 \to 0.531 &\qquad
    0.497 \to 0.699 &\qquad
    0.607 \to 0.747
  \end{pmatrix}
\end{align}
Note that there are strong correlations between the elements due to
the unitary constraint, see Ref.~\cite{GonzalezGarcia:2003qf} for
details on how we derive the ranges.

\begin{figure}\centering
  \includegraphics[width=0.9\textwidth]{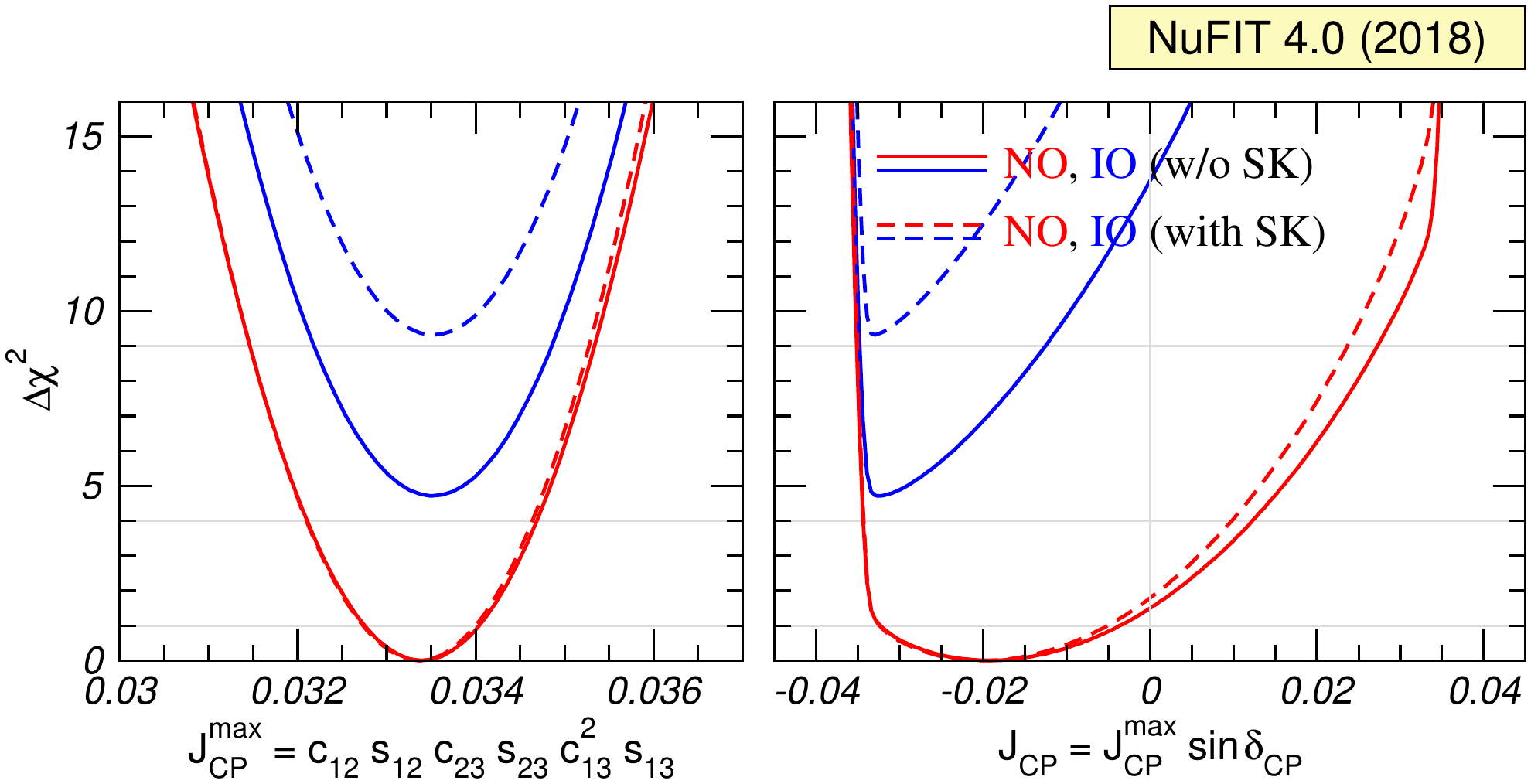}
  \caption{Dependence of the global $\Delta\chi^2$ function on the
    Jarlskog invariant. The red (blue) curves are for NO (IO). Solid
    (dashed) curves are without (with) adding the tabulated SK-atm
    $\Delta\chi^2$.}
  \label{fig:chisq-viola}
\end{figure}

The present status of leptonic CP violation is illustrated in
figs.~\ref{fig:region-glob} and~\ref{fig:chisq-viola}.  In particular
fig.~\ref{fig:region-glob} contains two projections of the confidence
regions with $\dCP$ on the vertical axis in which we observe the
non-trivial correlations between $\dCP$ and $\sin^2\theta_{23}$.  In
the left panel of fig.~\ref{fig:chisq-viola} we show the dependence of
$\Delta\chi^2$ of the global analysis on the Jarlskog invariant which
gives a convention-independent measure of CP
violation~\cite{Jarlskog:1985ht}, defined by:
\begin{align}
  J_\text{CP} &\equiv \Im\big[ U_{\alpha i} U_{\alpha j}^* U_{\beta i}^* U_{\beta j} \big]
  \nonumber\\
   &\equiv J_\text{CP}^\text{max} \sin\dCP =
  \cos\theta_{12} \sin\theta_{12}
  \cos\theta_{23} \sin\theta_{23} \cos^2\theta_{13} \sin\theta_{13}
  \sin\dCP
\end{align}
where in the second line we have used the parametrization in
Eq.~\eqref{eq:U3m}.  Factoring out $\sin\dCP$, the determination of
the mixing angles implies a maximal possible value of the Jarlskog
invariant:
\begin{equation}
  \label{eq:jmax}
  J_\text{CP}^\text{max} = 0.0333 \pm 0.0006 \, (\pm 0.0019)
\end{equation}
at $1\sigma$ ($3\sigma$) for both orderings.  The preference of the
present data for non-zero $\dCP$ implies a best fit value
$J_\text{CP}^\text{best} = -0.019$, which is favored over CP
conservation with $\Delta\chi^2 = 1.5 \,(1.8)$ without (with)
SK-atm. These numbers can be compared with the size of the Jarlskog
invariant in the quark sector, $J_\text{CP}^\text{quarks} = (3.18\pm
0.15) \times 10^{-5}$~\cite{PDG}.

\section{Synergies and tensions}
\label{sec:synergies}

\subsection{Status of comparison of results of solar experiments versus
  KamLAND}
\label{sec:synerg-dmq21}

\begin{figure}\centering
  \includegraphics[width=0.9\textwidth]{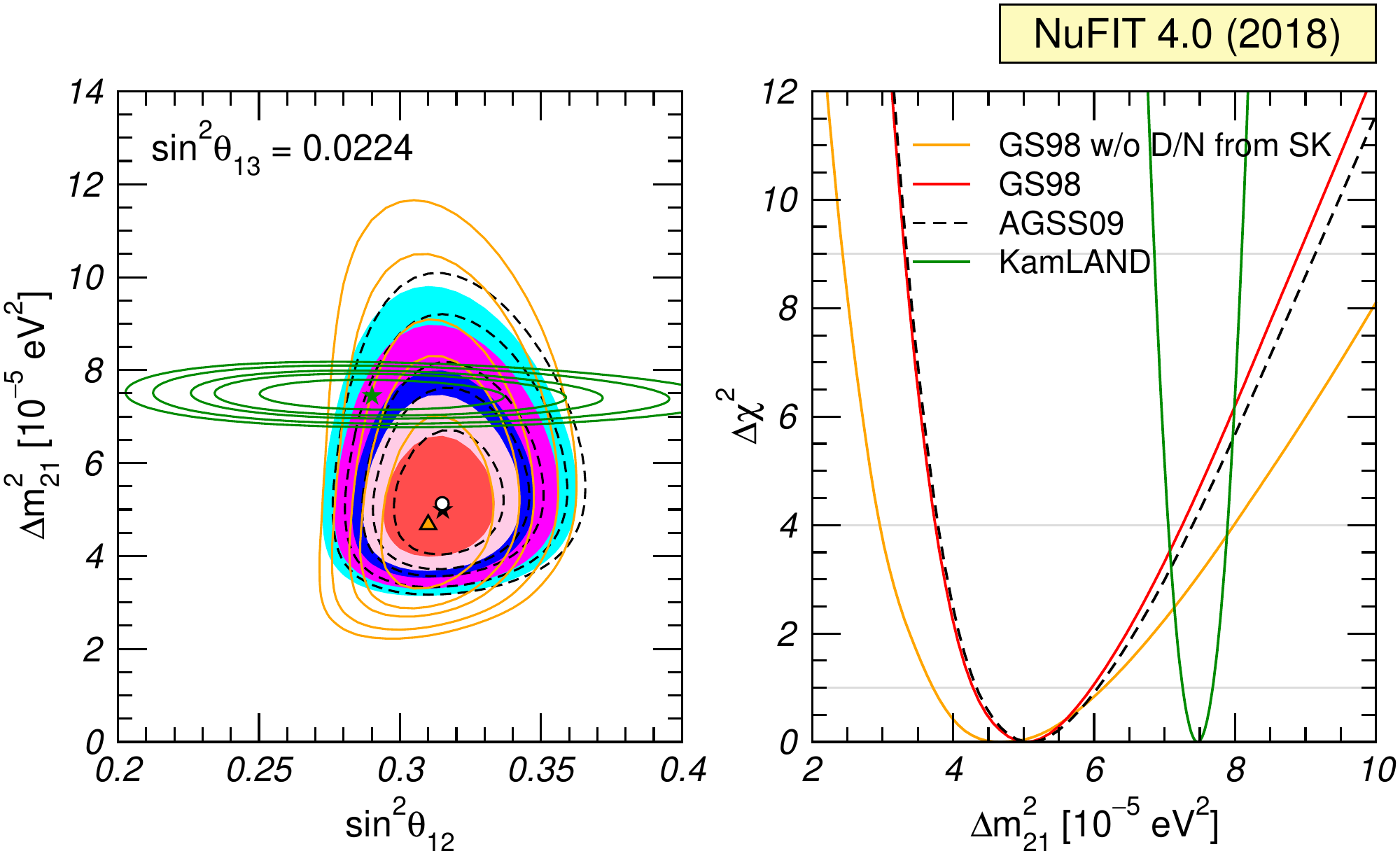}
  \caption{Left: Allowed parameter regions (at 1$\sigma$, 90\%,
    2$\sigma$, 99\%, and 3$\sigma$ CL for 2 dof) from the combined
    analysis of solar data for GS98 model (full regions with best fit
    marked by black star) and AGSS09 model (dashed void contours with
    best fit marked by a white dot), and for the analysis of KamLAND
    data (solid green contours with best fit marked by a green star)
    for fixed $\sin^2{\theta_{13}}=0.0224$ ($\theta_{13}=8.6$). We
    also show as orange contours the results of a global analysis for
    the GS98 model but without including the day-night information
    from SK. Right: $\Delta\chi^2$ dependence on $\Dmq_{21}$ for the
    same four analyses after marginalizing over $\theta_{12}$.}
  \label{fig:sun-tension}
\end{figure}

The analyses of the solar experiments and of KamLAND give the dominant
contribution to the determination of $\Dmq_{21}$ and $\theta_{12}$.
We show in fig.~\ref{fig:sun-tension} the present determination of
these parameters from the global solar analysis in comparison with
that of KamLAND data. The results of the solar neutrino analysis are
shown for the two latest versions of the Standard Solar Model, namely
the GS98 and the AGSS09 models~\cite{Bergstrom:2016cbh} obtained with
two different determinations of the solar
abundances~\cite{Vinyoles:2016djt}.  This clearly illustrates the
independence of the results with respect to the solar modeling.

There are two main differences compared to our previous published
results in Ref.~\cite{Esteban:2016qun}. In what respects the KamLAND
region it has shifted towards slightly smaller values of
$\theta_{12}$. This effect arises mainly from the new reactor fluxes
used in our analysis of the KamLAND data. As mentioned in
section~\ref{sec:data}, in our calculation of the event rates in KamLAND
we have replaced the predicted neutrino fluxes by the spectrum
measured in Daya Bay near detectors~\cite{An:2016srz} which is
unfolded for detector and remaining oscillation effects. In
Ref.~\cite{Esteban:2016qun} we used instead the unoscillated reactor
determined by including in the fit the results from a compilation of
short baseline reactor data. The net result is that the current
unoscillated reactor fluxes are slightly lower and consequently a
slightly higher survival probability is required to better fit the
data. Since in the context of 3$\nu$-oscillations
\begin{equation}
  P_{ee,\text{KLAND}}^{3\nu}
  = \sin^4\theta_{13} + \cos^4\theta_{13} \left(
  1 - \frac{1}{2}\sin^2(2\theta_{12}) \sin^2\frac{\Dmq_{21} L}{2E}
  \right)\,
\end{equation}
a larger survival probability implies smaller values of $\theta_{12}$.
As a result the best-fit value of $\theta_{12}$ determined by KamLAND,
$\sin^2\theta_{12,\text{bf-Kam}}=0.290$, does not perfectly align with
the corresponding best fit value from the solar neutrino analysis,
$\sin^2\theta_{12,\text{bf-sol}}=0.315$.  Statistically, however, this
is a very small effect as the best fit value of
$\sin^2\theta_{12}=0.315$ lies at $\Delta\chi^2_\text{KamLAND}\lesssim
1$.

In what respects the determination of $\Dmq_{21}$ it has been a result
of global analyses for several years already, that the value of
$\Dmq_{21}$ preferred by KamLAND is somewhat higher than the one from
solar experiments.  The tension arises from a combination of two
effects: the well-known fact that none of the \Nuc{8}{B} measurements
performed by SNO, SK and Borexino shows any evidence of the low energy
spectrum turn-up expected in the standard
LMA-MSW~\cite{Wolfenstein:1977ue, Mikheev:1986gs} solution for the
value of $\Dmq_{21}$ favored by KamLAND; and the observation of a
non-vanishing day-night asymmetry in SK, whose size is larger than the
one predicted for the $\Dmq_{21}$ value indicated of KamLAND.

The new addition to this issue in the present analysis is the
inclusion of the 2860-day energy spectrum of SK4~\cite{sksol:nu2018}
(compared to the 2365-day energy spectrum used in
~\cite{Esteban:2016qun}).  For the day-night variation of the results
we still use the SK4 2055-day day-night
asymmetry~\cite{sksol:nakano2016} because SK has not presented any
update concerning the day-night dependence of the observed rates. The
inclusion of the new spectral data makes the lack of the turn-up
effect slightly stronger (for example the best fit $\Dmq_{21}$ of
KamLAND was at $\Delta\chi^2_\text{solar}=4$ in the analysis
of~\cite{Esteban:2016qun} with the GS98 fluxes and it is now at
$\Delta\chi^2_\text{solar}=4.7$).  For illustration of the relevance
of the day-night variation results we plot in
fig.~\ref{fig:sun-tension} the corresponding results of the solar
analysis without including the day-night asymmetry.

\subsection{$\theta_{23}$, $\dCP$ and mass ordering from
  LBL accelerator and MBL reactor experiments}

\begin{figure}\centering
  \includegraphics[width=0.9\textwidth]{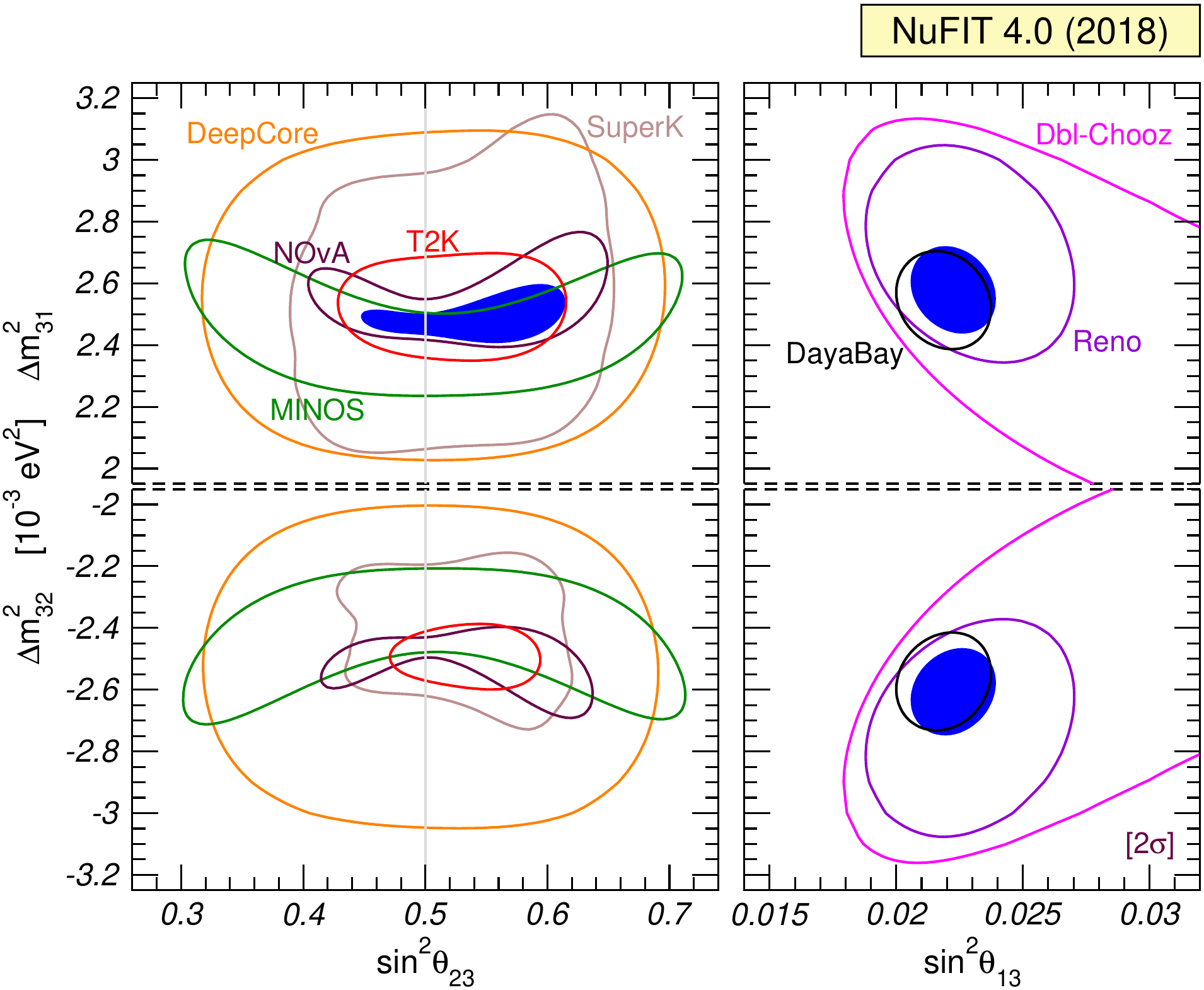}
  \caption{Determination of $\Dmq_{3\ell}$ at $2\sigma$ (2~dof), where
    $\ell=1$ for NO (upper panels) and $\ell=2$ for IO (lower
    panels). The left panels show regions in the $(\theta_{23},
    \Dmq_{3\ell})$ plane using both appearance and disappearance data
    from MINOS (green), T2K (red), NO$\nu$A (brown), as well as
    IceCube/DeepCore (orange), and SK-atm (from the table provided by
    the experiment, marron line) and the combination of them (blue
    coloured region).  In the left panels the constraint on
    $\theta_{13}$ from the global fit (which is dominated by the
    reactor data) is imposed as a Gaussian bias.  The right panels
    show regions in the $(\theta_{13}, \Dmq_{3\ell})$ plane using only
    Daya Bay (black), Reno (violet) and Double Chooz (magenta) reactor
    data, and their combination (blue coloured region). In all panels
    $\Dmq_{21}$, $\sin^2\theta_{12}$ are fixed to the global best fit
    values. Contours are defined with respect to the global minimum of
    the two orderings.}
  \label{fig:sample}
\end{figure}

The determination of the atmospheric parameters $\theta_{23}$ and
$\Dmq_{3\ell}$ is illustrated in fig.~\ref{fig:sample}. We observe
significant synergy from combining the various experiments, since the
combined region is clearly smaller than any individual one. Moreover,
the striking agreement of LBL accelerator and MBL reactor data in the
determination of $\Dmq_{3\ell}$ within comparable accuracy is a
non-trivial cross check of the 3-flavour oscillation paradigm.  Let us
now discuss in more detail how the indication of non-maximal mixing
and preference for the second octant for $\theta_{23}$ emerges.

\subsubsection{Disappearance results and non-maximal $\theta_{23}$}

\begin{figure}\centering
  \includegraphics[width=0.9\textwidth]{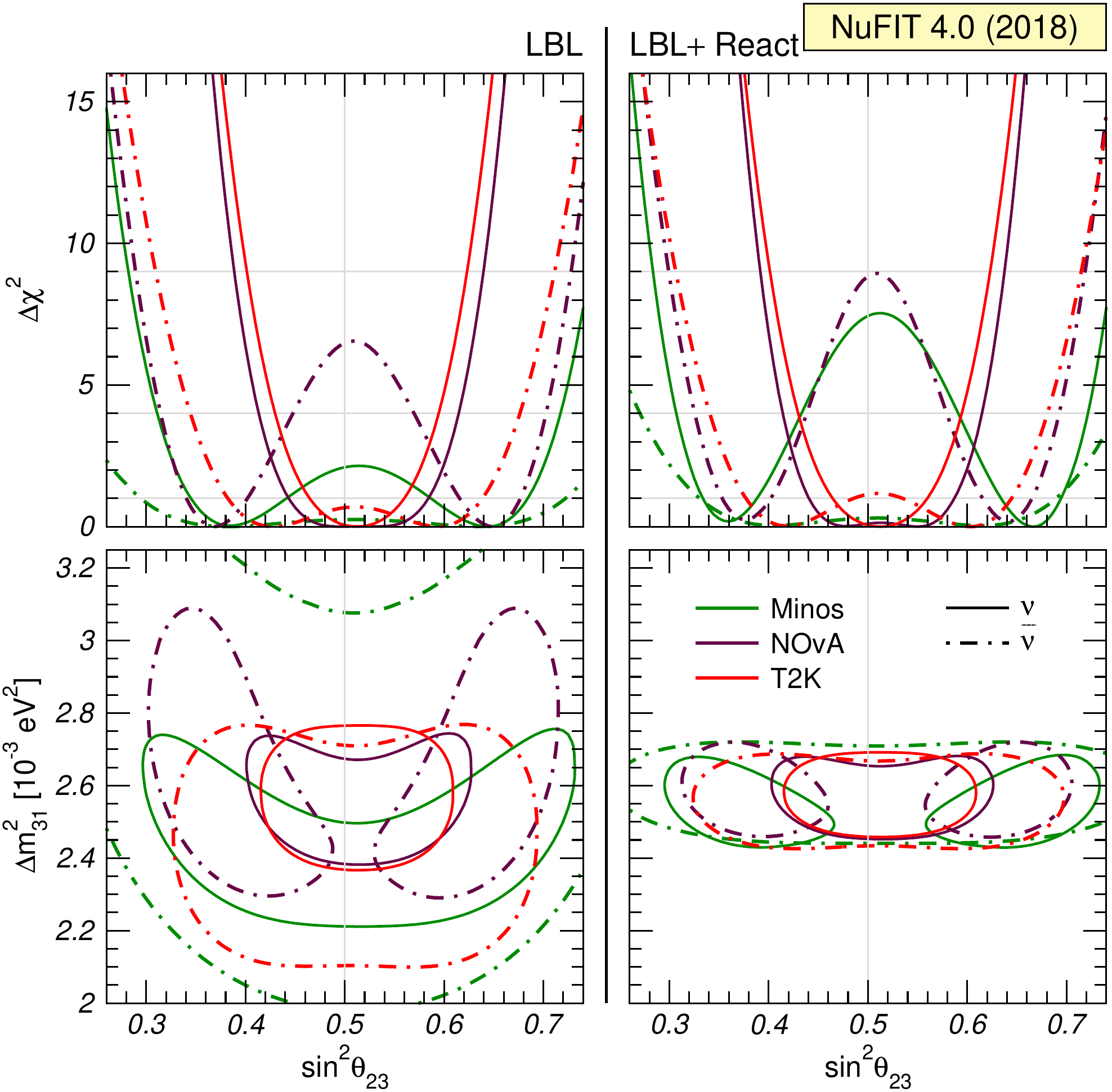}
  \caption{LBL accelerator $\nu_\mu$ disappearance data only, from
    MINOS, T2K, and NOvA, separated into neutrino and anti-neutrino
    data. Left panels correspond to LBL accelerator data with
    constraint on $\theta_{13}$ from the global fit (which is
    dominated by the MBL reactor data) imposed as a Gaussian bias. In
    the right panels LBL data are consistently combined with MBL
    reactor data from Daya Bay, RENO, and Double Chooz. Upper panels
    show the $\Delta\chi^2$ as a function of $\sin^2\theta_{23}$,
    lower panels show confidence regions at $2\sigma$ (2~dof). All
    panels assume NO and $\Dmq_{21}$, $\sin^2\theta_{12}$ are fixed to
    the global best fit values.  Qualitatively similar behaviour is
    found in IO.}
  \label{fig:dis}
\end{figure}

We focus first on LBL disappearance data. The $\nu_\mu$ survival
probability is given to good accuracy by~\cite{Okamura:2004if,
  Nunokawa:2005nx}
\begin{equation}
  P_{\mu\mu} \approx 1 - \sin^22\theta_{\mu\mu} \sin^2\frac{\Dmq_{\mu\mu} L}{4E_\nu} \,,
\end{equation}
where $L$ is the baseline, $E_\nu$ is the neutrino energy, and
\begin{align}
  \sin^2\theta_{\mu\mu} &= \cos^2\theta_{13} \sin^2\theta_{23} \,, \\
  \Dmq_{\mu\mu} &= \sin^2\theta_{12} \Dmq_{31} + \cos^2\theta_{12} \Dmq_{32} +
  \cos\dCP \sin\theta_{13}\sin 2\theta_{12} \tan\theta_{23} \Dmq_{21} \,.
  \label{eq:Dmqmm}
\end{align}
Hence the survival probability is symmetric with respect to the octant
of $\theta_{\mu\mu}$, which implies symmetry around $s_{23}^2 =
0.5/c_{13}^2 \approx 0.51$.  This behaviour is visible in the left
panels of fig.~\ref{fig:dis}, which show the results of LBL
accelerator disappearance data from MINOS, T2K, NOvA, separated into
the neutrino and anti-neutrino data samples (for fixed value of
$\theta_{13}$ at the best fit and NO).  While most of the shown data
samples prefer maximal mixing (especially T2K and NOvA neutrino data),
maximal mixing is disfavoured by MINOS neutrino data ($\Delta\chi^2
\approx 2$) and NOvA anti-neutrino data ($\Delta\chi^2 \approx 6$).
This behaviour can be traced back to the number of events in the
corresponding data samples observed at the dip of the survival
probability: for maximal mixing the survival probability is zero at
the dip and no events should be observed. Qualitatively similar
behaviour are found for IO.

In the lower-left panel of fig.~\ref{fig:dis} we observe in addition a
correlation between $\sin^2\theta_{23}$ and $\Dmq_{31}$ for the data
which prefer non-maximal mixing: larger values of $\Dmq_{31}$ imply
more deviation from maximal mixing. As visible in
fig.~\ref{fig:sample}, also MBL reactor data provide an accurate
determination of $\Dmq_{3\ell}$, which, however, pushes slightly to
larger values than LBL data. Because of the above mentioned
correlation, this leads to an even stronger preference for non-maximal
mixing, once LBL data are consistently combined with reactor data, as
visible in the right panels of fig.~\ref{fig:dis}: in combination with
reactors, MINOS neutrino and NOvA anti-neutrino data disfavour maximal
mixing with $\Delta\chi^2 \approx 7$ and 9, respectively.

\subsubsection{Appearance results, second $\theta_{23}$ octant and $\dCP$}

The preference for the second octant of $\theta_{23}$ is driven by
$\nu_\mu\to\nu_e$ appearance channel in LBL experiments (available
both for neutrinos and anti-neutrinos).  Following
Ref.~\cite{Elevant:2015ska}, the appearance probability can be
approximated by
\begin{align}
  P_{\nu_\mu\to\nu_e} & \approx 4 s_{13}^2s_{23}^2(1+2oA) - C \sin\dCP(1+oA) \,,\\
  P_{\bar\nu_\mu\to\bar\nu_e} &\approx 4 s_{13}^2s_{23}^2(1-2oA) + C \sin\dCP(1-oA) \,.
\end{align}
with
\begin{align}
  C \equiv \frac{\Dmq_{21}L}{4E_\nu}
  \sin2\theta_{12}\sin2\theta_{13}\sin2\theta_{23}
  \,,\quad
  o \equiv \text{sgn}(\Dmq_{3\ell}) \,,
  \quad A \equiv \left| \frac{2E_\nu V}{\Dmq_{3\ell}} \right| \,,
\end{align}
where $V$ is the effective matter potential. In the above equations we
have expanded in the small parameters $s_{13}$, $\Dmq_{21}L/E_\nu$,
and $A$, and used that for T2K and NOvA $|\Dmq_{3\ell}|L/4E_\nu
\approx \pi/2$.\footnote{Expanding in the matter potential parameter
  $A$ is a very good approximation for T2K, but not so good for
  NOvA. However, the qualitative behaviour is still captured by the
  above expressions also for NOvA, which suffices for our discussion
  here.}  Using the respective mean neutrino energies we find $A
\approx 0.05$ for T2K and an \emph{empirical} value of $A=0.1$ (for
which this approximation works better) at NOvA. Correspondingly the
number of observed appearance events in T2K and NOvA is approximately
proportional to the oscillation probability:
\begin{align}
  N_{\nu_e}
  &\approx \mathcal{N}_\nu \left[ 2 s_{23}^2(1+2oA) - C' \sin\dCP(1+oA) \right]\,,
  \label{eq:Nnu}
  \\
  N_{\bar\nu_e} &\approx \mathcal{N}_{\bar\nu}
  \left[ 2 s_{23}^2(1-2oA) + C' \sin\dCP(1-oA) \right]\,.
  \label{eq:Nan}
\end{align}
Taking all the well-determined parameters $\theta_{13}$,
$\theta_{12}$, $\Dmq_{21}$, $|\Dmq_{3\ell}|$ at their global best fit
points, we obtain numerically $C' \approx 0.28$. The normalization
constants $\mathcal{N}_{\nu,\bar\nu}$ calculated from our re-analysis
of T2K and NOvA are given for the various appearance samples in
table~\ref{tab:app}. Those values can be compared with the background
subtracted observed number of events, which we also report in the
table. Within this approximation, there are only the two parameters
$s_{23}^2$ and $\sin\dCP$, plus the discrete parameter $o=\pm 1$
encoding the mass ordering, to fit the appearance event numbers shown
in table~\ref{tab:app}, with $\sin^2\theta_{23}$ being constrained in
addition from disappearance data.  Note that $C'$ depends only on
$\sin2\theta_{23}$, which varies by less than 2\% for $0.42 < s_{23}^2
< 0.64$, and can be taken as constant for our purposes. The general
trends from eqs.~\eqref{eq:Nnu} and~\eqref{eq:Nan} are the following:
\begin{itemize}
\item Both neutrino and anti-neutrino events are enhanced by
  increasing $s_{23}^2$.

\item Values of $\sin\dCP \simeq +1\, (-1)$ suppress (increase)
  neutrino events, and have the opposite effect for anti-neutrino
  events.

\item For NO (IO) neutrino events are enhanced (suppressed) due to the
  matter effect, whereas anti-neutrino events are suppressed
  (enhanced).

\item For NO (IO) the matter effect increases (decreases) the impact
  of $\dCP$ for neutrinos, while the opposite happens for
  anti-neutrinos.
\end{itemize}
The last two items are more important for NOvA than for T2K, due to
larger matter effects in NOvA because of the longer baseline.

\begin{table}\centering
  \begin{footnotesize}
    \begin{tabular}{c|ccccc}
      \hline\hline
      & T2K CCQE ($\nu$) & T2K CC1$\pi$  ($\nu$) & T2K CCQE ($\bar\nu$) & NOvA ($\nu$) & NOvA ($\bar\nu$) \\
      \hline
      $\mathcal{N}$            & 40   & 3.8   & 11  & 34   & 11  \\
      $N_\text{obs}$             & 75   & 15    & 9   & 58   & 18 \\
      $N_\text{obs}-N_\text{bck}$  & 61.4 & 13.6  & 6.1 & 43.6 & 13.8 \\
      \hline\hline
    \end{tabular}
  \end{footnotesize}
  \caption{Normalization coefficients $\mathcal{N}_\nu$ and
    $\mathcal{N}_{\bar\nu}$ for eqs.~\eqref{eq:Nnu} and~\eqref{eq:Nan}
    for approximations used to qualitatively describe the various
    appearance event samples used in our analysis for T2K and NOvA.
    We also give the observed number of events, as well as the
    corresponding background subtracted event numbers, as reported in
    Refs.~\cite{wascko_morgan_2018_1286752,
      sanchez_mayly_2018_1286758}}
  \label{tab:app}
\end{table}

\begin{figure}\centering
  \includegraphics[width=0.9\textwidth]{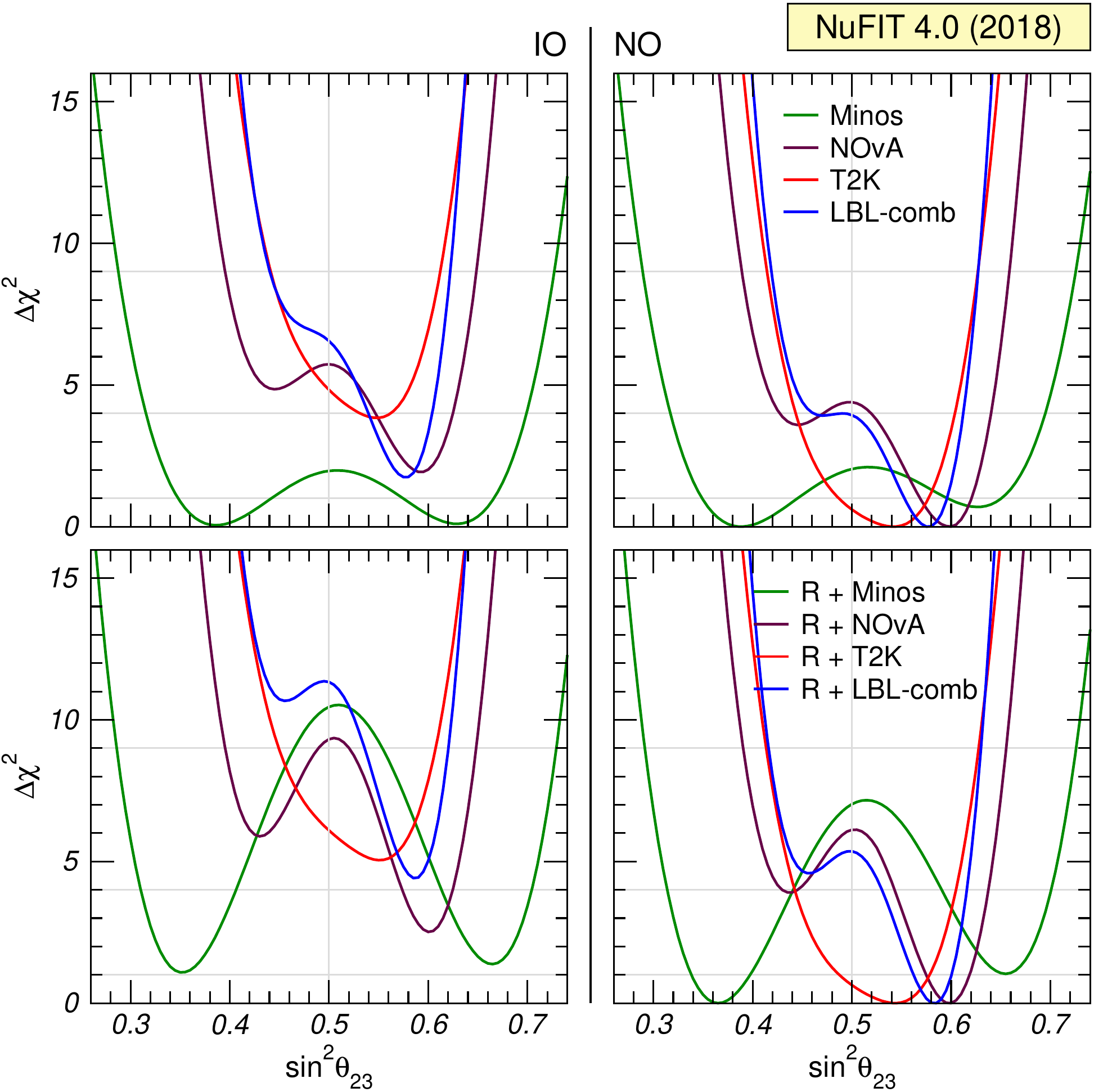}
  \caption{$\theta_{23}$ determination from LBL, reactor and their
    combination. Left (right) panels are for IO (NO). The upper panels
    show the 1-dim $\Delta\chi^2$ from LBL experiments after
    constraining \emph{only} $\theta_{13}$ from reactor
    experiments. For each experiment $\Delta\chi^2$ is defined with
    respect to the global minimum of the two orderings.  The lower
    panels show the corresponding determination when the full
    information of LBL accelerator and reactor experiments is used in
    the combination (including the information on $\Dmq_{3\ell}$ from
    reactors). In all panels $\Dmq_{21}$, $\sin^2\theta_{12}$ are
    fixed to the global best fit values.}
  \label{fig:t23}
\end{figure}

\begin{figure}\centering
  \includegraphics[width=0.9\textwidth]{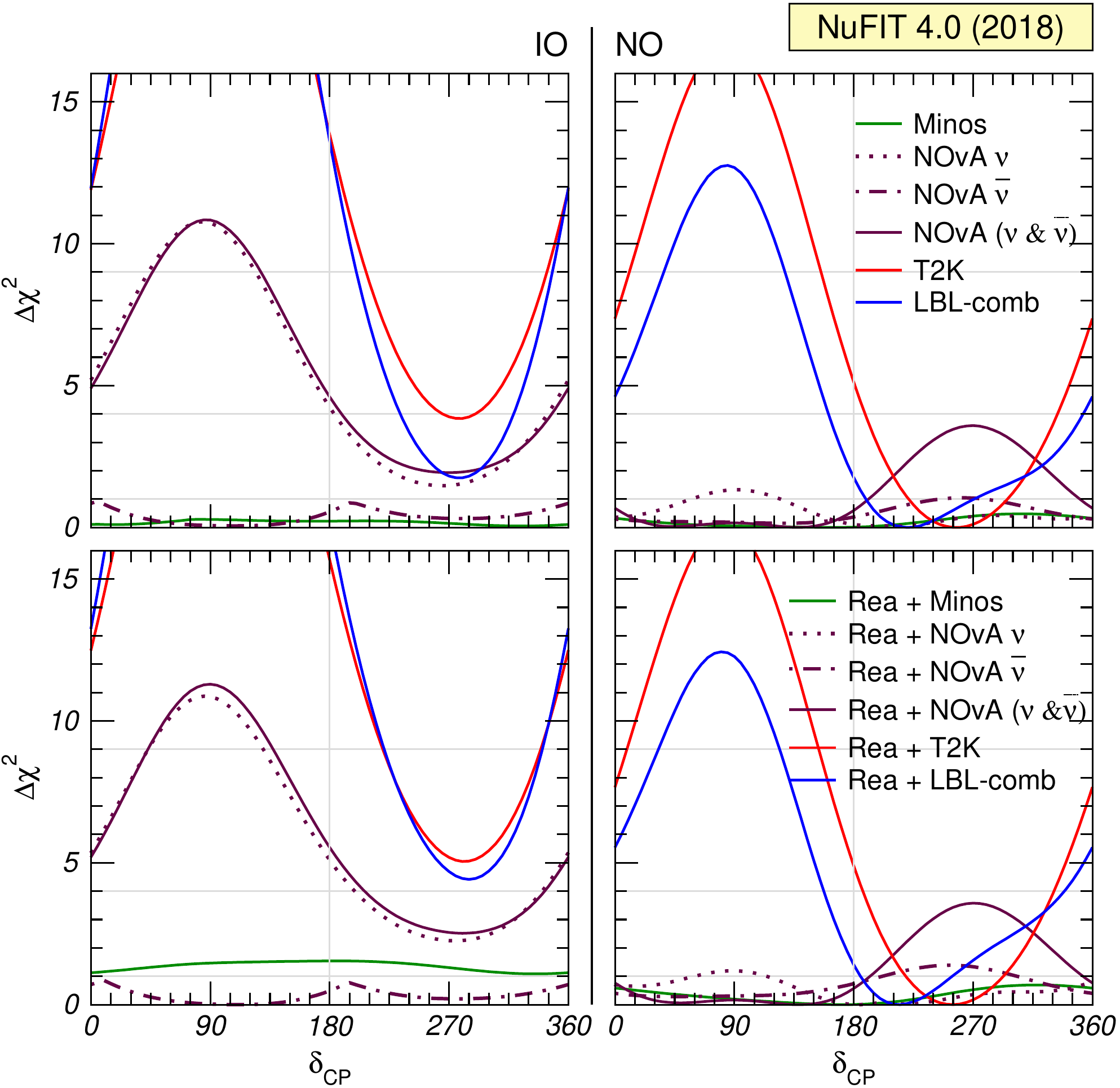}
  \caption{$\dCP$ determination from LBL, reactor and their
    combination. Left (right) panels are for IO (NO). The upper panels
    show the 1-dim $\Delta\chi^2$ from LBL experiments after
    constraining \emph{only} $\theta_{13}$ from reactor experiments.
    For each experiment $\Delta\chi^2$ is defined with respect to the
    global minimum of the two orderings.  The lower panels show the
    corresponding determination when the full information of LBL
    accelerator and reactor experiments on both mixing angles and
    $\Dmq_{3\ell}$ is used in the combination. In all panels
    $\Dmq_{21}$, $\sin^2\theta_{12}$ are fixed to the global best fit
    values.}
  \label{fig:dcp}
\end{figure}

In fig.~\ref{fig:t23}, the determination of $s_{23}^2$ from LBL data
(including appearance) combined with reactor data is shown. In the
upper panels only $\theta_{13}$ is constrained by reactor data,
whereas in the lower panels LBL and reactor data are combined
consistently, including also $\Dmq_{3\ell}$ information. For the
reasons explained above, lower panels show larger significance of
non-maximality, but now the symmetry between the octants is broken by
appearance data. Fig.~\ref{fig:dcp} shows the $\Delta\chi^2$
dependence on $\dCP$ for various data samples.

Let us consider first the T2K samples. We see from table~\ref{tab:app}
that in both neutrino samples (especially CC1$\pi$) the observed
number of events after background subtraction is large compared to
$\mathcal{N_\nu}$, while the anti-neutrino number is low. Hence, we
need to maximize the expression in eq.~\eqref{eq:Nnu} and minimize
eq.~\eqref{eq:Nan}. Since neutrino data dominates over anti-neutrinos,
a slight preference for $s_{23}^2 >0.5$ appears (constrained by
disappearance data), while at the same time $\sin\dCP \approx -1$
serves to maximize (minimize) neutrino (anti-neutrino) appearance, as
visible in fig.~\ref{fig:dcp}.

For NOvA neutrino data, the coefficient $\mathcal{N}_\nu$ in
eq.~\eqref{eq:Nnu} is also somewhat low compared to the observed
number of events minus background.  For NO, the matter effect enhances
neutrino events, and therefore, $s_{23}^2$ (around maximal mixing
favoured in disappearance) and $\dCP$ can be adjusted, such that the
event numbers can always be fitted, so $\Delta\chi^2(\dCP)$ from NOvA
neutrino data alone is $<1$ for NO, cf.~fig.~\ref{fig:dcp}. For IO,
however, the matter effect suppresses neutrino events, and therefore,
preference for the second octant and $\sin\dCP \approx -1$ appears to
maximize the term in the square-bracket in eq.~\eqref{eq:Nnu}.  For
NOvA anti-neutrino data, table~\ref{tab:app} shows that the observed
event number is of the order of $\mathcal{N}_{\bar\nu}$ (only slightly
higher).  Consequently we observe for NO only a very mild preference
for $\sin\dCP \approx 1$ just to enhance slightly the rate of
anti-neutrinos. For IO, the matter effect enhances anti-neutrinos, and
therefore, choosing the combinations (first $\theta_{23}$
octant/$\sin\dCP \approx 1$) or (second $\theta_{23}$ octant/$\sin\dCP
\approx -1$) can fit the events, which leads to negligible
$\Delta\chi^2(\dCP)$ dependence for IO NOvA anti-neutrinos,
cf.~fig.~\ref{fig:dcp}.  The combination of those effects for NO,
leads to a disfavouring of $\sin\dCP \approx -1$ with
$\Delta\chi^2\approx 3.5$ from NOvA, somewhat in contradiction of the
T2K preferred region: with the non-maximality of $\theta_{23}$ from
anti-neutrinos plus the matter enhancement for neutrinos, $\sin\dCP
\approx -1$ would predict too many neutrino events, and is therefore
disfavoured.

The conclusion of those considerations lead to the preference of the
second octant for $\theta_{23}$ in the global analysis, as well as
pushing the confidence interval for $\dCP$ towards $180^\circ$, which
implies that CP conservation is allowed by the combined data with
$\Delta\chi^2\approx 1.5$.

\subsubsection{Preference for normal ordering}

An important result of the present global fit is the growing
significance of the preference for the normal mass ordering.  This
indication emerges by a subtle interplay of various subsets of the
global data. Sensitivity to the mass ordering is provided by the
matter effect~\cite{Wolfenstein:1977ue, Barger:1980tf, Mikheev:1986gs}
in oscillations with $\Dmq_{3\ell}$, observable in LBL accelerator and
atmospheric neutrino experiments, as well as the comparison
oscillations in the $\nu_e$ and $\nu_\mu$ disappearance
channels~\cite{Nunokawa:2005nx, Minakata:2006gq, Blennow:2013vta}.

\begin{figure}\centering
  \includegraphics[width=0.9\textwidth]{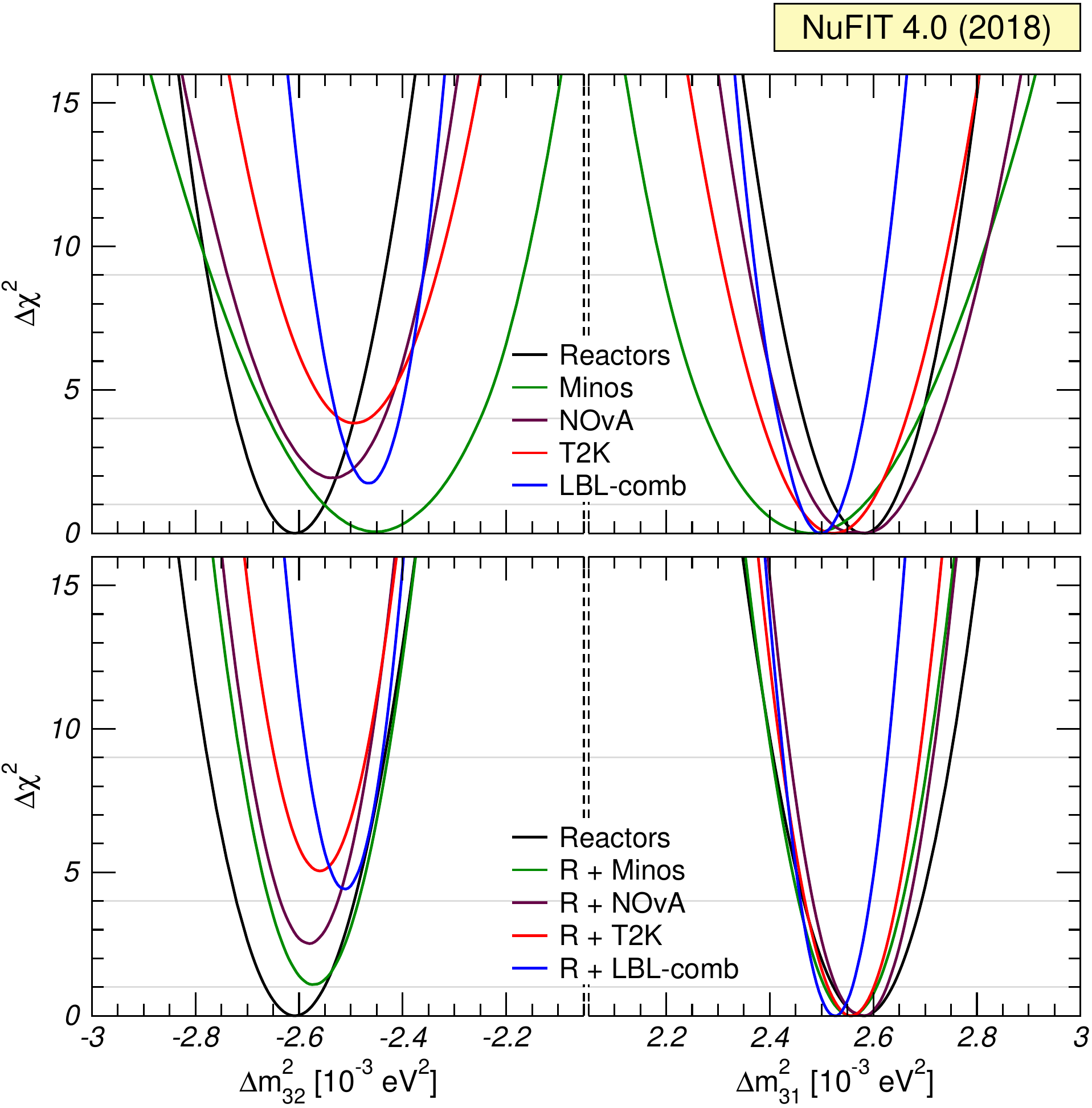}
  \caption{$\Dmq_{3\ell}$ determination from LBL, reactor and their
    combination. Left (right) panels are for IO (NO). The upper panels
    show the 1-dim $\Delta\chi^2$ from LBL experiments after
    constraining \emph{only} $\theta_{13}$ from reactor
    experiments. For each experiment $\Delta\chi^2$ is defined with
    respect to the global minimum of the two orderings.  The lower
    panels show the corresponding determination when the full
    information of LBL accelerator and reactor experiments is used in
    the combination (including the information on $\Dmq_{3\ell}$ from
    reactors). In all panels $\Dmq_{21}$, $\sin^2\theta_{12}$ are
    fixed to the global best fit values.}
  \label{fig:dma}
\end{figure}

Let us first discuss the indication coming from LBL accelerator
experiments.  We find that T2K + the $\theta_{13}$ constraint from
reactors disfavours IO by $\Delta\chi^2 \approx 4$, see upper panels
of figs.~\ref{fig:t23}, \ref{fig:dcp} and~\ref{fig:dma}. This can be
understood from the numbers in table~\ref{tab:app} and
eqs.~\eqref{eq:Nnu} and~\eqref{eq:Nan}, where the matter effect for NO
helps to increase (decrease) events for neutrinos
(anti-neutrinos). NOvA data + the $\theta_{13}$ constraint also
disfavours IO by about 2 units in $\chi^2$, driven by neutrino data,
while anti-neutrinos are insensitive to the ordering,
cf.~fig.~\ref{fig:dcp}. Interestingly, by combining T2K, NOvA, and
MINOS, \emph{decreases} the $\Delta\chi^2$ of IO to about 2. An
explanation for this effect is the slight tension between NOvA and T2K
in the determination of $\dCP$ for NO visible in
fig.~\ref{fig:dcp}. This leads to a worse fit of NO compared to IO,
where both experiments prefer the same region for $\dCP$.

An interesting additional effect sensitive to the mass ordering has
been pointed out in Refs.~\cite{Nunokawa:2005nx, Minakata:2006gq}: the
$\nu_\mu$ disappearance probability is symmetric with respect to the
sign of $\Dmq_{\mu\mu}$ given in eq.~\eqref{eq:Dmqmm}, while $\nu_e$
disappearance is symmetric with respect to a slightly different
effective mass-squared difference:
\begin{equation}
  \Dmq_{ee} = \cos^2\theta_{12} \Dmq_{31} + \sin^2\theta_{12} \Dmq_{32} \,.
  \label{eq:Dmqee}
\end{equation}
Hence, from a precise determination of the oscillation frequencies in
$\nu_\mu$ and $\nu_e$ disappearance experiments, information on the
sign of $\Dmq_{3\ell}$ can be obtained.\footnote{A similar effect has
  been exploited in Ref.~\cite{Blennow:2013vta}, based on the
  comparison of the $\Dmq_{3\ell}$ determination in future reactor and
  atmospheric neutrino experiments.} Indeed, we observe in
fig.~\ref{fig:dma} that this effect already contributes notably to the
mass ordering discrimination in present data: the upper panels show
the determination of $\Dmq_{3\ell}$ from the individual LBL
accelerator experiments ($\nu_\mu$ disappearance) compared to the one
from MBL reactors ($\nu_e$ disappearance). We have verified that those
curves are indeed symmetric with respect to the sign of
$\Dmq_{\mu\mu}$ and $\Dmq_{ee}$, respectively, within excellent
accuracy. When displaying them for common parameters ($\Dmq_{3\ell}$
in fig.~\ref{fig:dma}), we observe that the agreement is better for NO
than for IO. The difference between the upper and lower panels in the
$\Delta\chi^2$ for IO is largely due to this $\Dmq_{3\ell}$ effect. We
see that the $\Delta\chi^2$ for the LBL combination is pushed from 2
to about 4.5, when combined consistently with reactor data taking into
account the $\Dmq_{3\ell}$ dependence.

In summary, we obtain from LBL+reactor data a preference for NO at
about $2\sigma$. As mentioned in section~\ref{sec:oscparam}, this gets
further enhanced by atmospheric neutrino data, with the main
contribution from Super-Kamiokande, leading to the exclusion of IO at
about $3\sigma$, see fig.~\ref{fig:chisq-glob}. In the following
subsection we discuss in more detail various aspects of the
atmospheric neutrino analyses from IceCube and Super-Kamiokande.

\subsection{Treatment of atmospheric results from Super-Kamiokande and Deep-Core}
\label{sec:atm-data}

\begin{figure}\centering
  \includegraphics[width=0.9\textwidth]{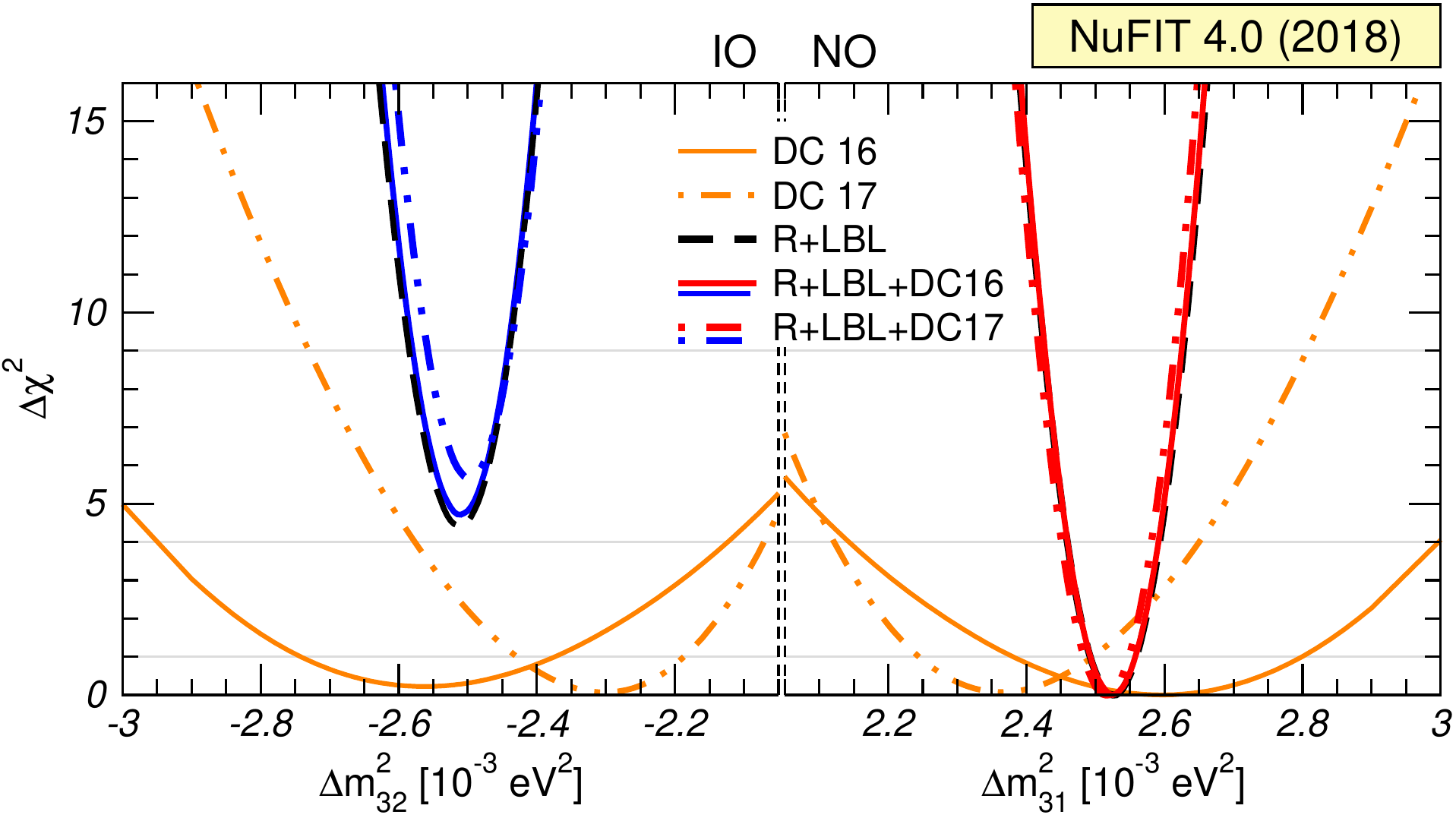}
  \caption{$\Delta\chi^2$ as a function of $\Dmq_{3\ell}$ for our
    reanalysis of Deep-Core 3-years data of
    Ref.~\cite{Aartsen:2014yll, deepcore:2016} (labeled DC16, solid
    orange line) and its combination with the global analysis of
    reactor and LBL experiments (full blue and red lines).  The
    corresponding dash-dotted line correspond to use the $\chi^2$
    table provided by the experiment for the analysis of their three
    years data in Ref.~\cite{Aartsen:2017nmd, deepcore:2017} (labeled
    DC17).  See text for details.}
  \label{fig:dc}
\end{figure}

In what respects the atmospheric neutrino data, in our default
analysis -- full lines in figs.~\ref{fig:chisq-glob}
and~\ref{fig:chisq-viola} (one-dimensional $\Delta\chi^2$ curves) and
coloured regions in fig.~\ref{fig:region-glob} (two-dimensional
projections of confidence regions) -- we include the results of the
Deep-Core 3-years data of Ref.~\cite{Aartsen:2014yll, deepcore:2016}
(which we refer here as DC16) for which the collaboration has provided
enough information on their effective areas to allow for our own
reanalysis. Its impact in the parameter determination obtained from
the combination of solar, reactor and LBL data is very marginal, see
fig.~\ref{fig:dc}, which displays as example its contribution to the
determination of $\Dmq_{3\ell}$ and the ordering where it adds about
0.3 units to $\chi^2_\text{min}$ of IO because of the slightly better
matching between the $\Dmq_{3\ell}$ from reactor+LBL experiments with
that of DC in NO.\footnote{All curves in fig.~\ref{fig:dc} contain the
  bias on $\Dmq_{21}$ and $\theta_{12}$ from solar and KamLAND, so the
  full lines denoted as R+LBL+DC16 coincide with the corresponding
  full lines in the corresponding panel in
  fig.~\ref{fig:chisq-glob}.}

In this respect it is interesting to notice that the ICECUBE
collaboration has recently published the results of a dedicated
analysis of another set of three-years data~\cite{Aartsen:2017nmd,
  deepcore:2017} leading to a better determination of $\Dmq_{3\ell}$
(which we refer to as DC17).  Unfortunately we cannot reproduce this
analysis because the corresponding effective areas have not been made
public.  The experiment has only made available the bi-dimensional
$\chi^2$ map (as a function of $\Dmq_{3\ell}$ and $\sin^2\theta_{23}$
for a fixed value of $\sin^2\theta_{13}=0.0217$ and $\dCP=0$)
corresponding to that analysis. Strictly this cannot be added in the
global analysis without making some assumption about their possible
$\theta_{13}$ and $\dCP$ dependence. Still, to illustrate the possible
impact of using these results we show also in fig.~\ref{fig:dc} the
corresponding contribution to the determination of $\Dmq_{3\ell}$ and
the ordering obtained by naively adding their $\chi^2$ map to our
results of the global reactor+LBL experiments (neglecting any possible
dependence on the fixed parameters). As seen in the figure, using the
DC17 results in the global combination disfavours IO by $\sim$ 1.2
additional units of $\chi^2$.  One must notice, however, that the
ICECUBE collaboration has recently performed a reanalysis of the same
data sample which leads to similar precision but a somewhat shifted
range for $\Dmq_{3\ell}$~\cite{deepcore:2017B}.

In what respects to the results of Super-Kamiokande, in the last five
years the collaboration has developed a more sophisticated analysis
method for their atmospheric neutrino data with the aim of
constructing $\nu_e + \bar\nu_e$ enriched samples which are then
further classified into $\nu_e$-like and $\bar\nu_e$-like subsamples,
thus increasing the sensitivity to subleading parameters such as the
mass ordering and $\dCP$. The official results obtained with this
method were published in Ref.~\cite{Abe:2017aap} and show -- once
$\theta_{13}$ is constrained to be within the range determined by
reactor experiments -- a preference for NO with a
$\Delta\chi^2(\text{IO}) = 4.3$, variation of $\chi^2(\dCP)$ with the
CP phase at the level of $\sim 90$\% CL (with favouring $\dCP\sim
270^\circ$), and a slight favouring of the second octant of
$\theta_{23}$ (see fig.14 in Ref.~\cite{Abe:2017aap}).

Unfortunately with the information at hand we are not able to
reproduce the elements driving the main dependence on these
subdominant oscillation effects in our own reanalysis of the data
samples which can be simulated outside of the collaboration.  However,
Super-Kamiokande has also published the results of their analysis in
the form of a tabulated $\chi^2$ map~\cite{SKatm:data2018} as a
function of the four relevant parameters $\Dmq_{3\ell},
\theta_{23},\theta_{13}$, and $\dCP$ which we can add to our global
analysis $\chi^2$ in the multidimensional parameter space in a fully
consistent form and then perform the corresponding parameter
marginalization to obtained the combined one-dimensional or two
dimensional parameter ranges.  The results of such combination are
shown as dashed curves in in figs.~\ref{fig:chisq-glob}
and~\ref{fig:chisq-viola} (one-dimensional $\Delta\chi^2$ curves) and
void regions in fig.~\ref{fig:region-glob} (two-dimensional
projections of confidence regions).  As can be seen from
fig.~\ref{fig:chisq-glob}, adding the SK-atm $\chi^2$ information
results into:
\begin{itemize}
\item Increase of the $\chi^2_\text{min}$ for IO by 4.6 units (from
  4.7 to 9.3).
\item Enhancement of the parameter dependence of $\chi^2(\dCP)$
  further disfavouring $\dCP$ values around $90^\circ$ (for example it
  increases $\chi^2(\dCP=70^\circ)$ in NO by $\sim$ 3 units from
  $\sim$ 13 to $\sim 16$)
\item Enhancement of the parameter dependence of $\chi^2(s^2_{23})$
  further disfavouring the first octant (for example it increases
  $\chi^2(s^2_{23}=0.45$--$0.5)$ in NO by $\sim$ 2 units.
\end{itemize}
In other words, as the SK-atm tendencies for these subdominant effects
are very well aligned with those from the combination of LBL
experiments (currently dominated by T2K), their impact in the
determination of $\dCP$ and $\theta_{23}$ in the global analysis is
almost equivalent to just adding for each of those parameters their
marginalized $\chi^2$ (with fixed $\theta_{13}$ at the reactor value)
to that from the global analysis without SK-atm.

\section{Projections on neutrino mass scale observables}
\label{sec:absmass}

\begin{figure}[t]\centering
  \includegraphics[width=0.9\textwidth]{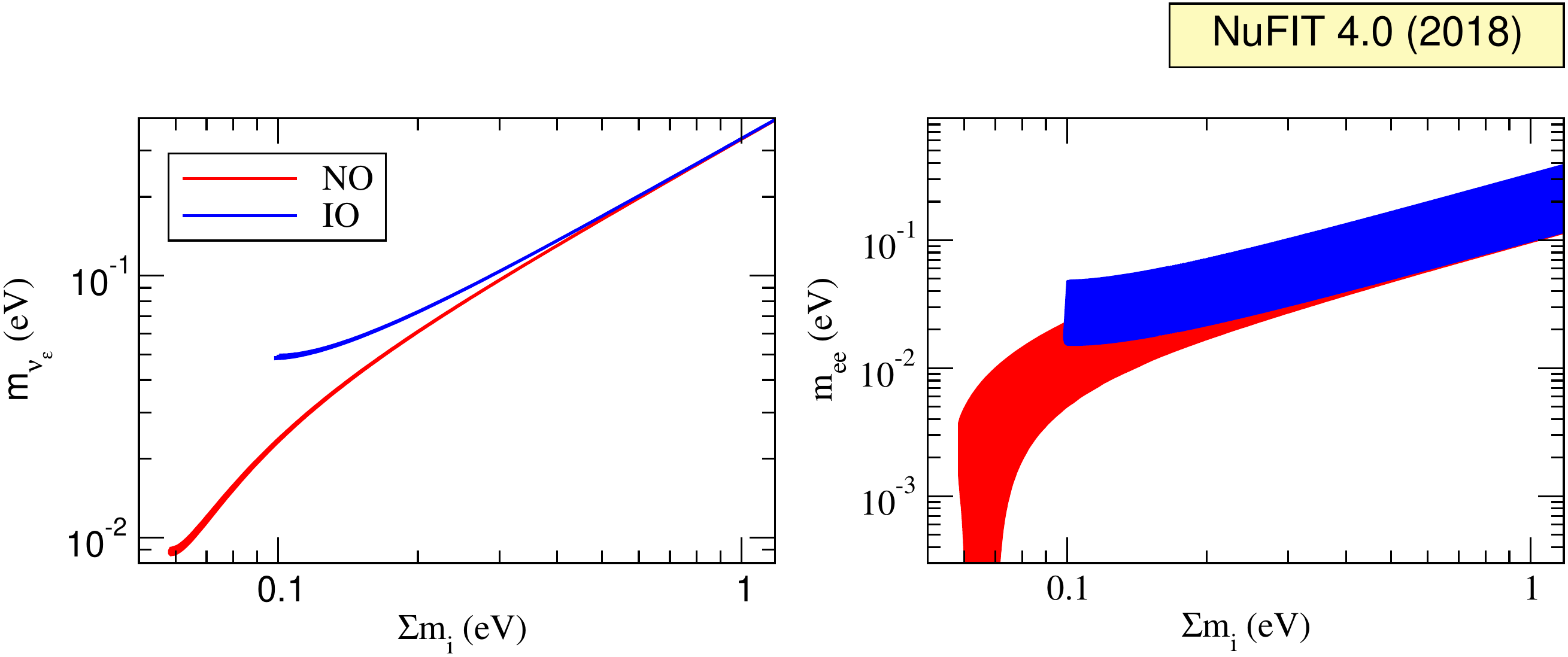}
  \caption{95\% allowed regions (for 2 dof) in the planes
    ($m_{\nu_e}$, $\sum m_\nu$) and ($m_{ee}$, $\sum m_\nu$) obtain
    from projecting the results of the global analysis of oscillation
    data. The regions are defined with respect to the minimum for each
    ordering.}
  \label{fig:3nu-mbeta}
\end{figure}

Oscillation experiments provide information on the mass-squared
splittings $\Dmq_{ij}$ and on the leptonic mixing angles $U_{ij}$, but
they are insensitive to the absolute mass scale for the neutrinos.  Of
course, the results of an oscillation experiment do provide a lower
bound on the heavier mass in $\Dmq_{ij}$, $|m_i| \geq
\sqrt{\Dmq_{ij}}$ for $\Dmq_{ij} > 0$, but there is no upper bound on
this mass. In particular, the corresponding neutrinos could be
approximately degenerate at a mass scale that is much higher than
$\sqrt{\Dmq_{ij}}$.  Moreover, there is neither an upper nor a lower
bound on the lighter mass $m_j$.

Information on the neutrino masses, rather than mass differences, can
be extracted from kinematic studies of reactions in which a neutrino
or an anti-neutrino is involved. In the presence of mixing the most
relevant constraint comes from the study of the end point ($E \sim
E_0$) of the electron spectrum in Tritium beta decay $\Nuc{3}{H} \to
\Nuc{3}{He} + e^- + \bar\nu_e$. This spectrum can be effectively
described by a single parameter, $m_{\nu_e}$, if for all neutrino
states $E_0 - E \gg m_i$:
\begin{equation}
  \label{eq:mbeta}
  \begin{aligned}
    m^2_{\nu_e}
    &= \frac{\sum_i m^2_i |U_{ei}|^2}{\sum_i |U_{ei}|^2}
    = \sum_i m^2_i |U_{ei}|^2
    = c_{13}^2 c_{12}^2 m_1^2
    + c_{13}^2 s_{12}^2 m_2^2+s_{13}^2 m_3^2
    \\
    &= \begin{cases}
      \text{NO: }
      & m^2_0 + \Dmq_{21} c_{13}^2 s_{12}^2+\Dmq_{3\ell} s_{13}^2 \,,
      \\
      \text{IO: }
      & m^2_0 - \Dmq_{21} c_{13}^2 c_{12}^2-\Dmq_{3\ell} c_{13}^2
    \end{cases}
  \end{aligned}
\end{equation}
where the second equality holds if unitarity is assumed and $m_0=m_1\,
(m_3)$ in NO (IO) denotes the lightest neutrino mass.  At present we
only have an upper bound, $m_{\nu_e} \leq 2.2$~eV at 95\%
CL~\cite{Bonn:2001tw}, which is expected to be superseded soon by
KATRIN~\cite{Osipowicz:2001sq} with about one order of magnitude
improvement in sensitivity.

Direct information on neutrino masses can also be obtained from
neutrinoless double beta decay $(A,Z) \to (A,Z+2) + e^- + e^-$.  This
process violates lepton number by two units, hence in order to induce
the $0\nu\beta\beta$ decay, neutrinos must be Majorana particles. In
particular, for the case in which the only effective lepton number
violation at low energies is induced by the Majorana mass term for the
neutrinos, the rate of $0\nu\beta\beta$ decay is proportional to the
\emph{effective Majorana mass of $\nu_e$}:
\begin{multline}
  m_{ee} = \Big| \sum_i m_i U_{ei}^2 \Big|
  = \Big| m_1 c_{13}^2 c_{12}^2 e^{i 2\alpha_1} +
  m_2 c_{13}^2 s_{12}^2 e^{i 2\alpha_2} +
  m_3 s_{13}^2 e^{-i 2\dCP} \Big|
  \\
  = \begin{cases}
    \text{NO:}
    & m_0\, \Big|c_{13}^2 c_{12}^2 e^{i 2(\alpha_1-\dCP)}
    + \sqrt{1 + \frac{\Dmq_{21}}{m_0^2}} \, c_{13}^2 s_{12}^2 e^{i 2(\alpha_2-\dCP)}
    + \sqrt{1 + \frac{\Dmq_{3\ell}}{m_0^2}}\, s_{13}^2 \Big|
    \\
    \text{IO:}
    & m_0\, \Big| \sqrt{1-\frac{\Dmq_{3\ell} +\Dmq_{21}}{m_0^2}} \,
    c_{13}^2 c_{12}^2 e^{i 2(\alpha_1 - \dCP)}
    + \sqrt{1-\frac{\Dmq_{3\ell}}{m_0^2}}\,
    c_{13}^2 s_{12}^2 e^{i 2(\alpha_2-\dCP)} + s_{13}^2 \Big|
  \end{cases}
\end{multline}
which, unlike Eq.~\eqref{eq:mbeta}, depends also on the CP violating
phases.
Recent searches have established the lifetime of this decay to be
longer than $\sim$ $10^{26}$ yr~\cite{Agostini:2018tnm,
  KamLAND-Zen:2016pfg}, corresponding to a limit on the neutrino mass
of $m_{ee} \lesssim 0.06-0.200$ eV at 90\% CL.  A series of new
experiments is planned with sensitivity of up to $m_{ee} \sim
0.01$~eV~\cite{GomezCadenas:2011it}.

Neutrino masses have also interesting cosmological effects. In
general, cosmological data mostly give information on the sum of the
neutrino masses, $\sum_i m_{i}$, while they have very little to say on
their mixing structure and on the ordering of the mass states.

Correlated information on these three probes of the neutrino mass
scale can be obtained by mapping the results from the global analysis
of oscillations presented previously. We show in
fig.~\ref{fig:3nu-mbeta} the present status of this exercise.  The
relatively large width of the regions in the right panel are due to
the unknown Majorana phases. Thus from a positive determination of two
of these probes, in principle information can be obtained on the value
of the Majorana phases and/or the mass ordering~\cite{Fogli:2004as,
  Pascoli:2005zb}.

\section{Conclusions}
\label{sec:conclu}

We have presented the results of the updated (as of fall 2018)
analysis of relevant neutrino data in the framework of mixing among
three massive neutrinos. We have shown our results for two analyses.
The first contains our own statistical combination of all the
experimental data for which we are able to reproduce the results of
the partial analysis performed by the different experiments, and
therefore does not include the information of the Super-Kamiokande
atmospheric neutrino data. In the second analysis we combine the
likelihood of the first one with the four-dimensional $\chi^2$ map
provided by Super-Kamiokande for the analysis of their atmospheric
data.  Quantitatively the present determination of the two mass
differences, three mixing angles and the relevant CP violating phase
for the two analysis is listed in table~\ref{tab:bfranges}, and the
corresponding leptonic mixing matrix is given in
Eq.~\eqref{eq:umatrix}. In both analysis the maximum allowed CP
violation in the leptonic sector parametrized by the Jarlskog
determinant is $J_\text{CP}^\text{max} = 0.0333 \pm 0.0006 \, (\pm
0.0019)$ at $1\sigma$ ($3\sigma$).

We have performed a detail study of the role of the different data
samples and their correct combination in the determination of the less
known parameters, $\theta_{23}$, $\dCP$ and the ordering in
section~\ref{sec:synergies}.  We can summarize the main conclusions in
this section as follows:
\begin{itemize}
\item The long standing tension between the best $\Dmq_{21}$
  determined in the solar neutrino analysis and that from KamLAND
  persists.  The inclusion of latest spectral data from SK4 and the
  use of Daya Bay near detector data for reactor flux normalization in
  KamLAND has made this tension slightly stronger, but it is still a
  $\sim 2\sigma$ effect.

\item We obtain preference for the second octant of $\theta_{23}$ in
  the global analysis with a best fit at $\sin^2\theta_{23} =
  0.58$. There are two effects contributing to this results:
  \begin{itemize}
  \item While most data samples in $\nu_\mu$ and $\bar\nu_\mu$
    disappearance at LBL prefer close to maximal mixing (especially
    T2K and NOvA neutrino data), maximal mixing is disfavoured by
    MINOS neutrino data ($\Delta\chi^2 \approx 2$) and NOvA
    anti-neutrino data ($\Delta\chi^2 \approx 6$). This disfavouring
    increases when fully combining with the reactor neutrino
    determination of $\Dmq_{3\ell}$ to $\Delta\chi^2 \approx 7$ and 9,
    respectively.

  \item The appearance results both in T2K and NOvA (SK-atm adds in
    the same direction) favour the second octant.  The final value of
    the best fit $\theta_{23}$ results of this effect in combination
    with the substantial non-maximality favoured by NOvA anti-neutrino
    and MINOS neutrino disappearance data.
  \end{itemize}

\item The determination of $\dCP$ is mostly driven by T2K neutrino and
  anti-neutrino appearance results which favour $\dCP\sim 3\pi/2$ and
  disfavours $\dCP\sim \pi/2$ for both NO and IO. NOvA neutrino
  appearance data align with this behaviour, and are more
  statistically significant in IO. On the contrary NOvA anti-neutrino
  appearance data are better (worse) described with $\dCP\sim \pi/2$
  ($\dCP\sim 3\pi/2$) in NO.  This slight tension results into a shift
  of the best fit to $\dCP = 215^\circ$ in NO. So the allowed range is
  pushed towards the CP conserving value of $180^\circ$, which now is
  only disfavoured with $\Delta\chi^2 \lesssim 2$.
  
\item Regarding the mass ordering:
  \begin{itemize}
 \item Both T2K and NOvA prefer NO individually: T2K (NOvA) + the
   $\theta_{13}$ constraint from reactors disfavours IO by
   $\Delta\chi^2 \approx 4\, (2)$, but combining T2K, NOvA (and MINOS)
   \emph{decreases} the $\Delta\chi^2$ of IO to about 2. This is a
   consequence of the slight tension between NOvA and T2K in the
   determination of $\dCP$ for NO.

 \item Additional sensitivity to the ordering is found by the precise
   determination of the oscillation frequency in $\nu_\mu$ and $\nu_e$
   disappearance data at LBL and reactors, respectively.  This effect
   increases $\Delta\chi^2$ of IO from 2 to about 4.5.
   
 \item Inclusion of the atmospheric neutrino results (mainly from SK)
   further increases $\Delta\chi^2$ of IO to the 3$\sigma$ level.
  \end{itemize}
\end{itemize}
Future updates of this analysis will be provided at the NuFIT website
quoted in Ref.~\cite{nufit}.

\section*{Acknowledgments}

This work is supported by USA-NSF grant PHY-1620628, by EU Networks
FP10 ITN ELUSIVES (H2020-MSCA-ITN-2015-674896) and INVISIBLES-PLUS
(H2020-MSCA-RISE-2015-690575), by MINECO grant FPA2016-76005-C2-1-P
and MINECO/FEDER-UE grants FPA2015-65929-P and FPA2016-78645-P, by
Maria de Maetzu program grant MDM-2014-0367 of ICCUB, and by the
``Severo Ochoa'' program grant SEV-2016-0597 of
IFT. I.E.\ acknowledges support from the FPU program fellowship
FPU15/03697.

\appendix

\section{List of data used in the analysis}
\label{sec:appendix-data}

\subsection*{Solar experiments}

\begin{itemize}
\item \emph{External information}: Standard Solar
  Model~\cite{Vinyoles:2016djt}.

\item Chlorine total rate~\cite{Cleveland:1998nv}, 1 data point.

\item Gallex \& GNO total rates~\cite{Kaether:2010ag}, 2 data points.

\item SAGE total rate~\cite{Abdurashitov:2009tn}, 1 data point.

\item SK1 full energy and zenith spectrum~\cite{Hosaka:2005um}, 44
  data points.

\item SK2 full energy and day/night spectrum~\cite{Cravens:2008aa}, 33
  data points.

\item SK3 full energy and day/night spectrum~\cite{Abe:2010hy}, 42
  data points.

\item SK4 2055-day day-night asymmetry~\cite{sksol:nakano2016} and
  2860-day energy spectrum~\cite{sksol:nu2018}, 24 data points.

\item SNO combined analysis~\cite{Aharmim:2011vm}, 7 data points.

\item Borexino Phase-I 741-day low-energy data~\cite{Bellini:2011rx},
  33 data points.

\item Borexino Phase-I 246-day high-energy data~\cite{Bellini:2008mr},
  6 data points.

\item Borexino Phase-II 408-day low-energy
  data~\cite{Bellini:2014uqa}, 42 data points.
\end{itemize}

\subsection*{Atmospheric experiments}

\begin{itemize}
\item \emph{External information}: Atmospheric neutrino
  fluxes~\cite{Honda:2015fha}.

\item IceCube/DeepCore 3-year data~\cite{Aartsen:2014yll,
  deepcore:2016}, 64 data points.

\item
  SK1-4 328 kiloton years~\cite{Abe:2017aap}, $\chi^2$ map~\cite{SKatm:data2018}
  added to our global analysis.
\end{itemize}

\subsection*{Reactor experiments}

\begin{itemize}
\item KamLAND separate DS1, DS2, DS3 spectra~\cite{Gando:2013nba} with
  Daya Bay reactor $\nu$ fluxes~\cite{An:2016srz}, 69 data points.

\item Double Chooz FD-I/ND and FD-II/ND spectral ratios, with 455-day
  (FD-I), 363-day (FD-II) and 258-day (ND)
  exposures~\cite{dc:cabrera2016}, 56 data points.

\item Daya Bay 1958-day EH2/EH1 and EH3/EH1 spectral
  ratios~\cite{Adey:2018zwh}, 52 data points.

\item Reno 2200-day FD/ND spectral ratios~\cite{Bak:2018ydk}, 26 data points.
\end{itemize}

\subsection*{Accelerator experiments}

\begin{itemize}
\item MINOS $10.71\times 10^{20}$~pot $\nu_\mu$-disappearance
  data~\cite{Adamson:2013whj}, 39 data points.

\item MINOS $3.36\times 10^{20}$~pot $\bar\nu_\mu$-disappearance
  data~\cite{Adamson:2013whj}, 14 data points.

\item MINOS $10.6\times 10^{20}$~pot $\nu_e$-appearance
  data~\cite{Adamson:2013ue}, 5 data points.

\item MINOS $3.3\times 10^{20}$~pot $\bar\nu_e$-appearance
  data~\cite{Adamson:2013ue}, 5 data points.

\item T2K $14.93\times 10^{20}$ pot $\nu_\mu$-disappearance
  data~\cite{t2k:vietnam2016}, 55 data points.

\item T2K $14.93\times 10^{20}$ pot $\nu_e$-appearance
  data~\cite{t2k:vietnam2016}, 23 data points for the CCQE and 16 data
  points for the CC1$\pi$ samples.

\item T2K $11.24\times 10^{20}$ pot $\bar\nu_\mu$-disappearance
  data~\cite{t2k:koga2018}, 55 data points.

\item T2K $11.24\times 10^{20}$ pot $\bar\nu_e$-appearance
  data~\cite{t2k:koga2018}, 23 data points.

\item NO$\nu$A $8.85\times 10^{20}$ pot $\nu_\mu$-disappearance
  data~\cite{sanchez_mayly_2018_1286758}, 76 data points.

\item NO$\nu$A $8.85\times 10^{20}$ pot $\nu_e$-appearance
  data~\cite{sanchez_mayly_2018_1286758}, 13 data points.

\item NO$\nu$A $6.91\times 10^{20}$ pot $\bar{\nu}_\mu$-disappearance
  data~\cite{sanchez_mayly_2018_1286758}, 76 data points.

\item NO$\nu$A $6.91\times 10^{20}$ pot $\bar{\nu}_e$-appearance
  data~\cite{sanchez_mayly_2018_1286758}, 12 data points.
\end{itemize}

\section{Technical details and validation cross checks}
\label{sec:appendix-checks}

This this appendix we provide some details on our analysis of the most
recent data from accelerator and reactor experiments, and show that we
can reproduce the results of the experimental collaborations with good
accuracy, when using the same assumptions in the analysis.

\subsection{T2K}

The predicted number of events in the T2K far detector in a given
energy bin $i$ and for a given channel $\alpha$ can be calculated as
\begin{equation}
  N_i^\alpha = N_\text{bkg,i}
  + \int^{E_{i+1}}_{E_i} \, \mathrm{d}E_\text{rec} \int_0^\infty \,
  \mathrm{d}E_\nu R(E_\text{rec}, E_\nu) \frac{\mathrm{d} \Phi}{\mathrm{d} E_\nu}
  \sigma_\alpha (E_\nu) \varepsilon (E_\nu) P_{\nu_\mu \to \nu_\alpha}(E_\nu) \,,
  \label{eq:T2K-nevts}
\end{equation}
where
\begin{itemize}
\item $N_\text{bkg,i}$ is the number of background events in that bin,
  which we have extracted from Ref.~\cite{t2k:vietnam2016} and
  consistently re-scaled to the latest exposure. If there is a
  neutrino component, its oscillation has to be consistently included.

\item $[E_i, E_{i+1}]$ are the bin limits.

\item $E_\text{rec}$ is the reconstructed neutrino energy.
  
\item $E_\nu$ is the true neutrino energy.
  
\item $R(E_\text{rec}, E_\nu)$ is the energy reconstruction function,
  that we take to be Gaussian.

\item $\frac{\mathrm{d} \Phi}{\mathrm{d} E_\nu }$ is the incident
  $\nu_\mu$ flux, extracted from Ref.~\cite{Dennis:2015cfe}.

\item $\sigma_\alpha$ is the $\nu_\alpha$ detector cross-section,
  extracted from Ref.~\cite{Dennis:2015cfe}.

\item $\varepsilon$ is the detection efficiency, which is adjusted to
  reproduce the observed spectra in Ref.~\cite{t2k:koga2018}.

\item $P_{\nu_\mu \to \nu_\alpha} (E_\nu)$ is the $\nu_\mu \to
  \nu_\alpha$ oscillation probability.
\end{itemize}
For the antineutrino channel, one has to switch $\nu$ by $\bar{\nu}$.

\begin{figure}\centering
  \includegraphics[width=0.9\textwidth]{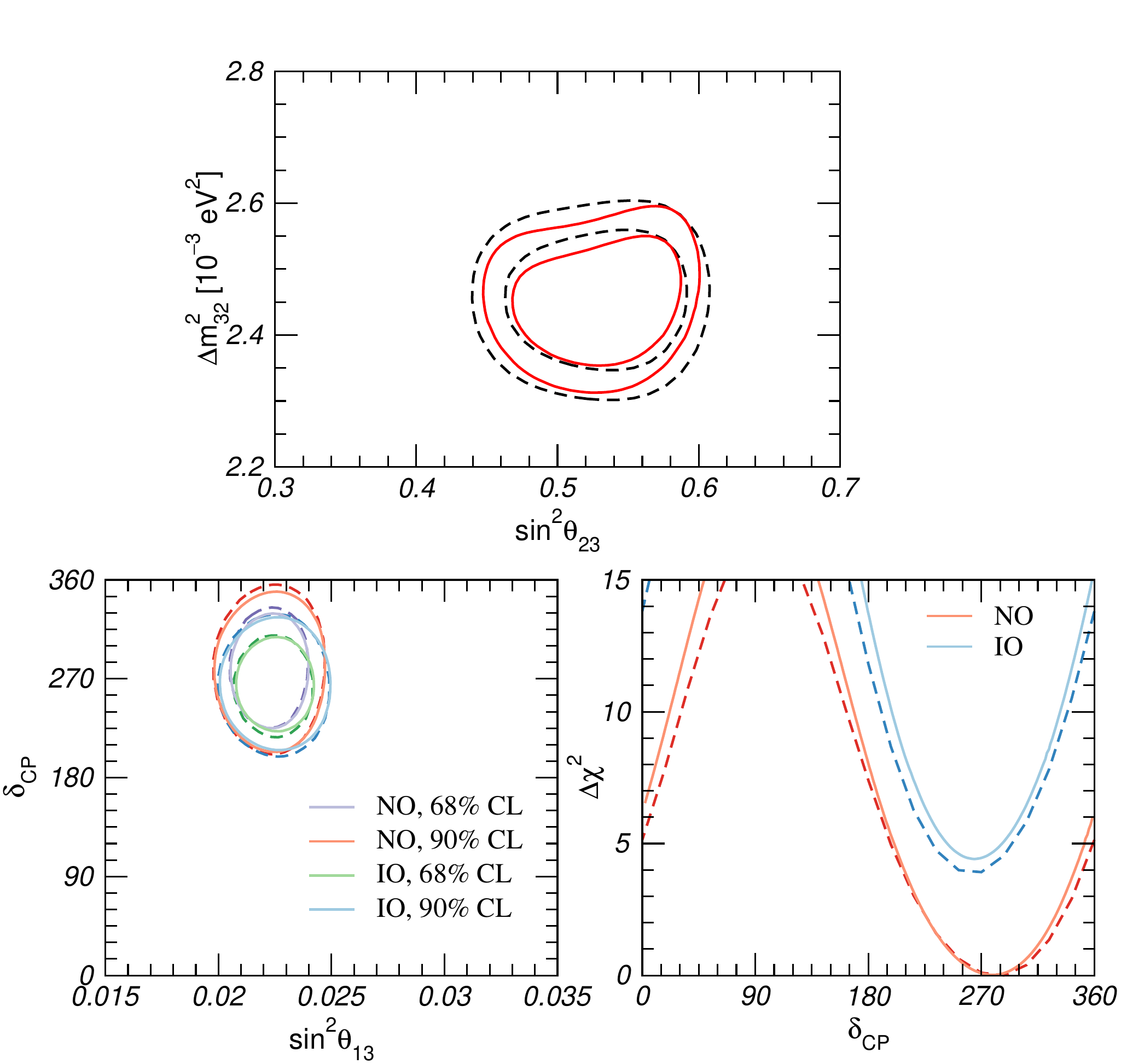}
  \caption{Different projections to our fit to latest T2K data (dashed
    lines) compared to the corresponding results of the experimental
    collaboration (solid curves), as presented in
    Ref.~\cite{t2k:koga2018}, when adopting the same assumptions. All
    unshown parameters are marginalised over.}
  \label{fig:t2k-check}
\end{figure}

If we assume a Poissonian $\chi^2$ with the data points in
Ref.~\cite{t2k:koga2018}, add an overall normalisation systematic
uncertainty\footnote{We take it to be 7\% for the $\nu_\mu$ and
  $\bar{\nu}_\mu$ disappearance channels, 5\% for the $\bar{\nu}_e$
  appearance channel, 8\% for the CCQE $\nu_e$ appearance channel, and
  23\% for the CC1$\pi$ $\nu_e$ appearance channel.} and combine all
the data, we get the contours in fig.~\ref{fig:t2k-check}. The other
oscillation parameters are fixed as specified in
Ref.~\cite{t2k:koga2018} and the reactor uncertainty on $\theta_{13}$
is included as a Gaussian bias and marginalised over. Finally,
following the ad-hoc procedure described in section~8.4.2 in
Ref.~\cite{t2k:koga2018}, the disappearance $\Dmq_{32}$ contours are
manually Gaussian-smeared with a standard deviation $\sigma = 4.1
\cdot 10^{-5}~\eVq$.

\subsection{NOvA}

\begin{figure}\centering
  \includegraphics[width=0.9\textwidth]{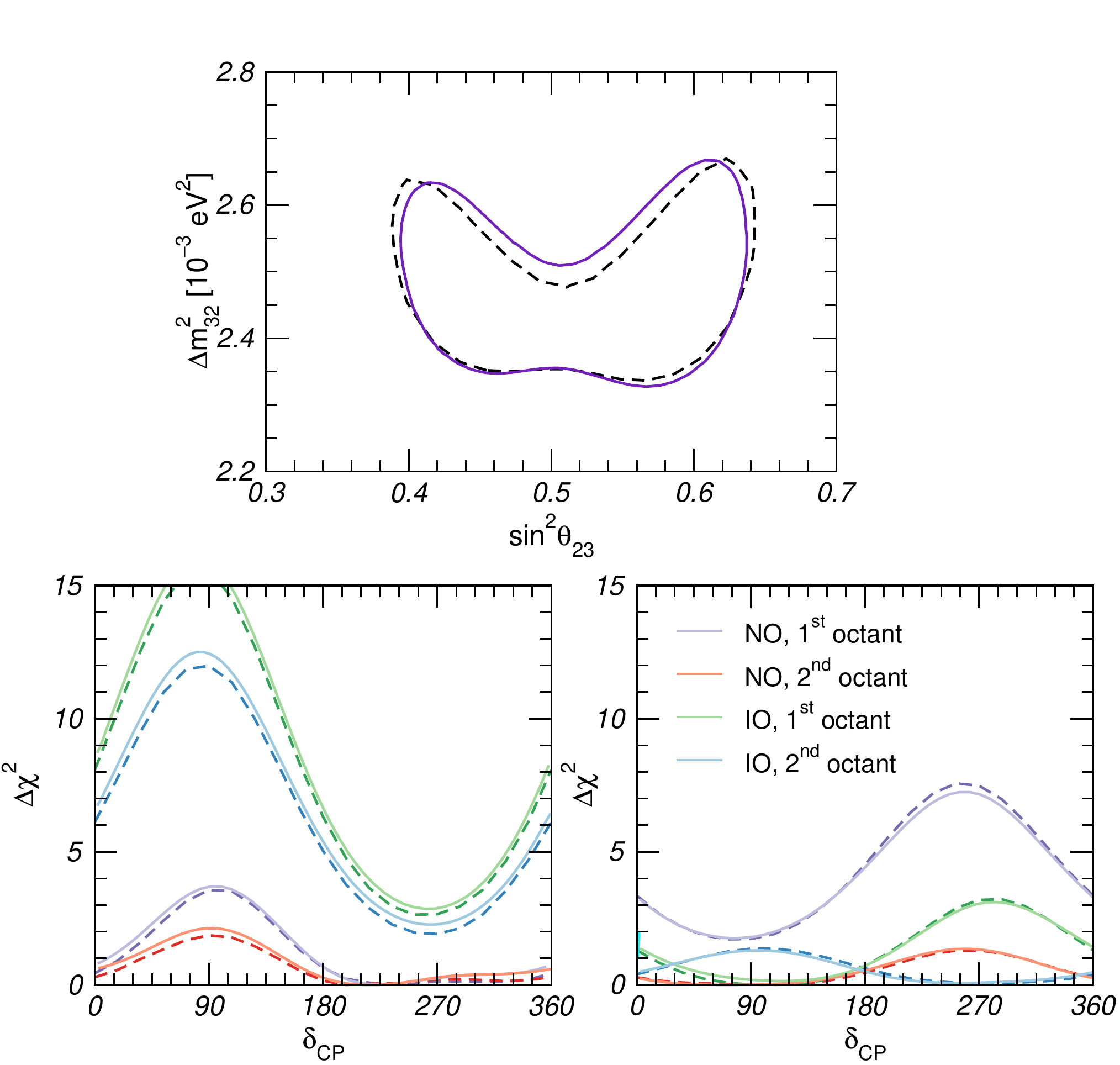}
  \caption{Different projections to our fit to latest NOvA data
    (dashed lines) compared to the corresponding results of the
    experimental collaboration (solid curves), as presented in
    Ref.~\cite{sanchez_mayly_2018_1286758}, when adopting the same
    assumptions. The upper plot only includes disappearance data,
    whereas the bottom left plot corresponds to $\nu_e$ appearance and
    the bottom right plot to $\bar{\nu}_e$ appearance.}
  \label{fig:nova-check}
\end{figure}

For NOvA the predicted number of events is given, as for T2K, by
Eq.~\eqref{eq:T2K-nevts}. Fluxes and detector response are extracted
from Refs.~\cite{baird_michael_2018_1301120, nova:WC2018}. We take the
data points and backgrounds from Ref.~\cite{nova:WC2018} and with
those construct a Poissonian $\chi^2$ including an normalisation
systematic uncertainty\footnote{We take it to be 5\% for $\nu_e$
  appearance, 6\% for $\bar{\nu}_e$ appearance, and 6\% for both the
  $\nu_\mu$ and $\bar{\nu}_\mu$ disappearance channels. We take these
  uncertainties as fully correlated for the disappearance chanels.}
and we get the contours in fig.~\ref{fig:nova-check} when the
undisplayed oscillation parameters are fixed as specified in
Ref.~\cite{nova:WC2018}. In particular the reactor uncertainty on
$\theta_{13}$ is included as a Gaussian bias and marginalised over.

\subsection{Daya Bay}
\label{sec:appendix-DB}

The Daya Bay $3\nu$ oscillation analysis is based
on~\cite{Adey:2018zwh}, the data is taken from the supplementary
material provided on arXiv.

In each experimental hall in Daya Bay there are more than one
detector, so the number of events in hall $H$, $N^H_i$ is a sum of the
contributions in all the detectors in the hall.  The predicted numbers
of events in a detector $d$ in an energy bin $i$ is computed as
\begin{equation}
  N_i^d = \mathcal{N} \sum_r \sum_{\text{iso}}
  \frac{\epsilon^d}{L^2_{rd}} \int^{E^{rec}_{i+1}}_{E^{\text{rec}}_i}
  dE^{\text{rec}} \int^{\infty}_{0} dE_{\nu} \,
  \sigma(E_{\nu}) f^{\text{iso}} \phi^{\text{iso}}(E_{\nu})
  P^{rd}_{\bar \nu_e \to \bar \nu_e}(E_{\nu}) R(E^{\text{rec}},E_{\nu}) \,.
  \label{eq:number-of-events}
\end{equation}
The indices $i$, $r$, $d$, iso refer to the energy bin, reactor,
detector, and fissible isotope, respectively.  $\epsilon^d$ are the
detector efficiencies, including $\varepsilon_\mu \times
\varepsilon_m$ multiplied by the life time days (taken both from
table~I in~\cite{Adey:2018zwh}) as well as the relative difference of
target protons $\Delta N_p$ in each detector, obtained from table~VI
in~\cite{An:2016ses}.  $L_{rd}$ are the baselines between the reactor
$r$ and detector $d$, obtained from table~I in~\cite{An:2016ses}.
$E_{\nu}$ and $E^{\text{rec}}$ are the true and reconstructed neutrino
energy, which are related by the detector response function
$R(E^{\text{rec}},E_\text{prompt})$, provided in the complementary
material to~\cite{Adey:2018zwh}. The relation between the prompt
energy $E_\text{prompt}$ and the neutrino energy $E_{\nu}$ is $E_{\nu}
= m_n - m_p - m_e + E_\text{prompt}$.  $\sigma(E_{\nu})$ is the
Inverse Beta Decay cross section computed performing the integral over
$\cos \theta$ of the differential cross section
in~\cite{Vogel:1999zy}. $\phi^{\text{iso}}(E_{\nu})$ are the
Huber-Mueller flux predictions~\cite{Huber:2011wv, Mueller:2011nm} and
$f^{\text{iso}}$ are the fission fractions.  For each isotope,
$f^{\text{iso}}$ is computed as the average of the fission fractions
in table~9 of Ref.~\cite{An:2016srz}.
Following section~2.6 of~\cite{An:2016srz}, we apply non-equilibrium
corrections by adding the relative correction from table~VII
of~\cite{Mueller:2011nm} to the fluxes~\cite{Huber:2011wv,
  Mueller:2011nm}.  Since Daya Bay has run for a long period we take
the row corresponding to 450 days. $P^{rd}_{\bar \nu_e \to \bar
  \nu_e}(E_{\nu})$ is the oscillation probability. The global constant
$\mathcal{N}$ will cancel when taking ratios of event numbers.

\begin{figure}\centering
  \includegraphics[width=0.49\textwidth]{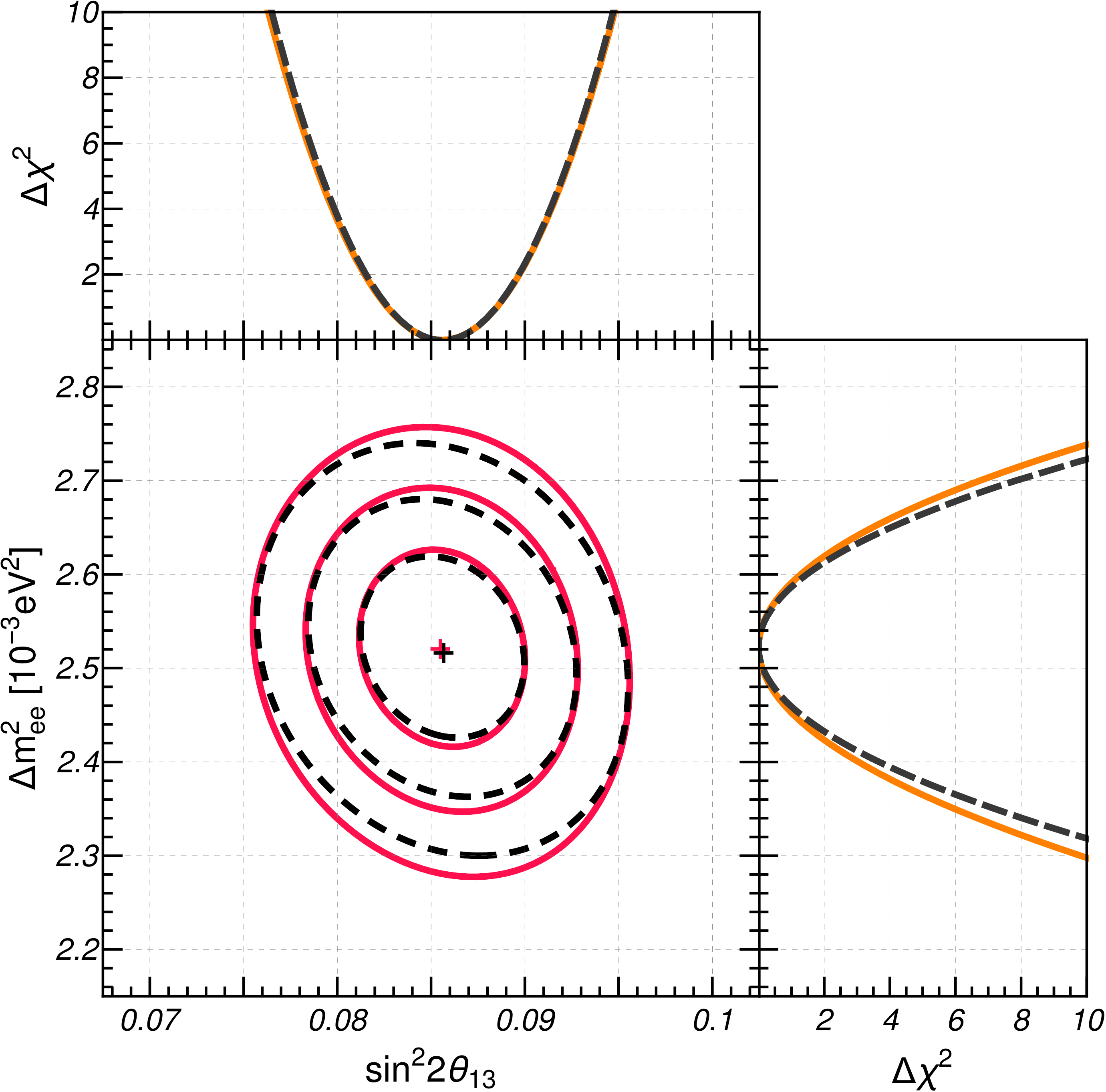}
  \includegraphics[width=0.49\textwidth]{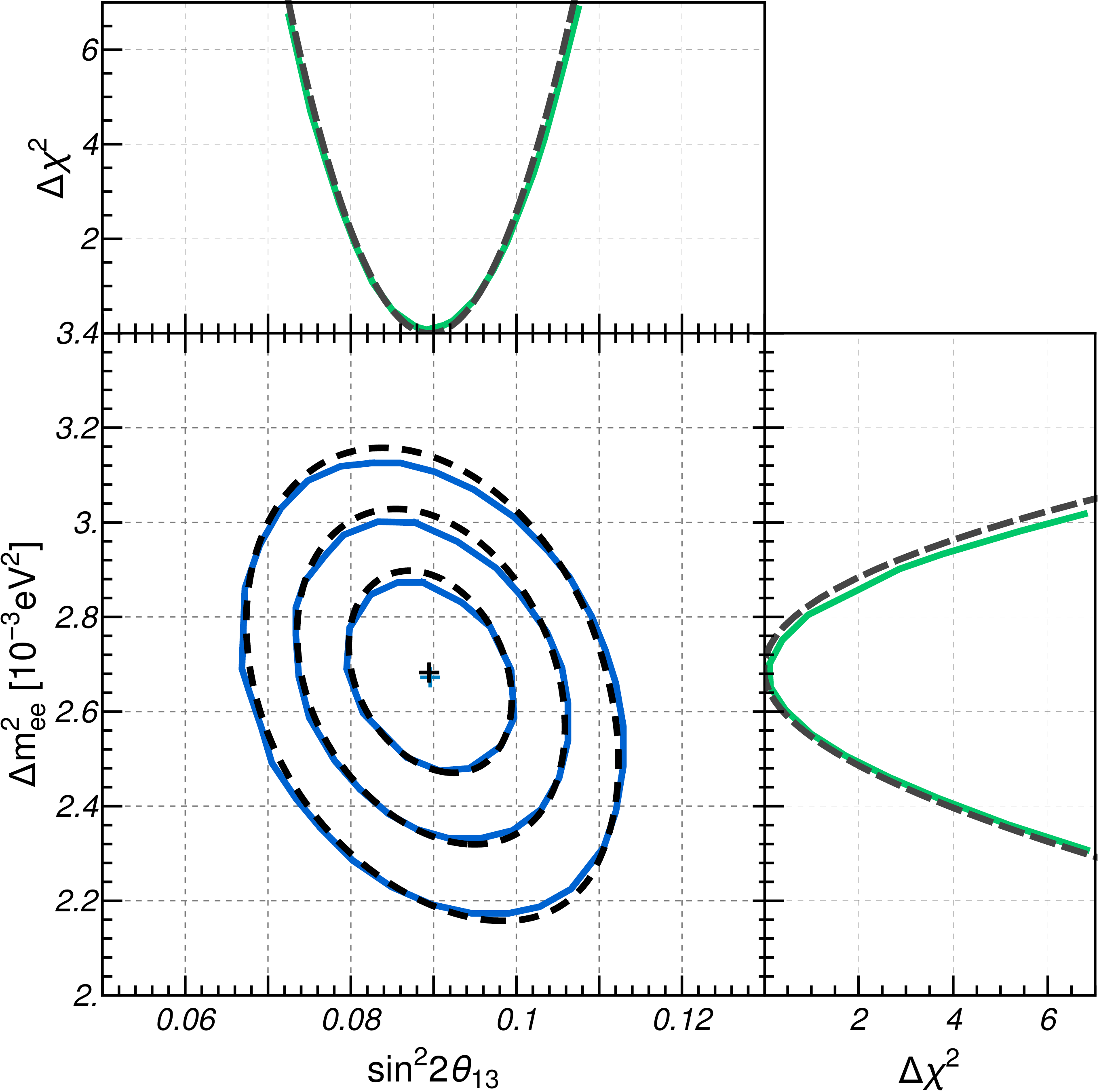}
  \caption{Our fit to DayaBay and RENO (black-dashed lines) compared
    to the results of the experimental collaborations (solid lines),
    as published in Refs.~\cite{Adey:2018zwh} and~\cite{Bak:2018ydk},
    respectively, when adopting the same assumptions. Following the
    collaborations, on the vertical axes, the parameter $\Dmq_{ee}$
    defined in eq.~\eqref{eq:Dmqee} is used.}
  \label{fig:reactor-check}
\end{figure}

Our DayaBay $\chi^2$ is based on the ratios of the observed spectra in
experimental halls 3 and 1 as well as 2 and 1:
\begin{multline}
  \label{eq:DBchisq}
  \chi^2(\theta_{12},\theta_{13},\Dmq_{21},\Dmq_{31}, \vec\eta) =
  \sum_i \frac{\left( \frac{O^3_i-B^3_i(\vec\eta)}{O^1_i-B^1_i(\vec\eta)}
    - \frac{N^3_i}{N^1_i}(\theta_{12},\theta_{13},\Dmq_{21},\Dmq_{31}, \vec\eta) \right)^2}{\left(\sigma^{stat_{31} }_i \right)^2}
  \\
  + \sum_i \frac{\left( \frac{O^2_i-B^2_i(\vec\eta)}{O^1_i-B^1_i(\vec\eta)} - \frac{N^2_i}{N^1_i}(\theta_{12},\theta_{13},\Dmq_{21},\Dmq_{31}, \vec\eta) \right)^2}
  {\left(\sigma^{stat_{21}}_i\right)^2}
  + \vec\eta^T V_\eta^{-1} \vec\eta \,. 
\end{multline}
Here, $O^{H}_i$ and $B^{H}_i(\vec\eta)$ are the observed number of
events and the background predictions in the experimental hall $H$ and
bin $i$, which can be found in the complementary material
of~\cite{Adey:2018zwh}. $\sigma^{stat_{H H'}}_i$ are the errors of the
ratio $[O^{H}_i - B^{H}_i(\vec\eta)] \,\big/\, [O^{H'}_i -
  B^{H'}_i(\vec\eta)]$, computed propagating the statistical errors of
$O^{H}_i$ and $O^{H'}_i$.\footnote{We neglect the correlation of the
  statistical errors due to the events in EH1, which appear in both
  ratios. In this way we obtain better agreement with the results of
  the DayaBay collaboration, cf.~fig.~\ref{fig:reactor-check},
  indicating that this choice of errors provides a fair
  approximation.}

In eq.~\eqref{eq:DBchisq}, $\vec\eta$ is the vector of pull parameters
and $V_\eta$ is the pull correlation matrix which accounts for the
systematic uncertainties and their correlations.  In order to
reproduce the Daya Bay results~\cite{Adey:2018zwh}, systematics in
detection efficiency, relative energy scale and cosmogenic Li-He
background have to be taken into account.  The detection efficiency
and relative energy scale uncertainties are given by 0.13\% and 0.2\%,
respectively~\cite{Adey:2018zwh, An:2016ses}, and the Li-He background
uncertainty is given by 30\%~\cite{Adey:2018zwh}.  We include also the
uncertainties in accidental (1\%) and fast neutron (13\% (17\%) in EH1
and EH2 (EH3)) backgrounds~\cite{An:2016ses}, which however play only
a subleading role. We take into account the 5 systematics in each of
the 3 experimental hall as uncorrelated, so $V_\eta$ is a diagonal
$15\times 15$ matrix.  In order to use the correct uncertainties in
each experimental hall we divide the detector uncertainties by $\sqrt
2$ in EH1 and EH2 and by $\sqrt{4}$ in EH3, since there are 2 and 4
detectors, respectively

In the left panel of fig.~\ref{fig:reactor-check}, our re-analysis is
compared to the one published in~\cite{Adey:2018zwh}.

\subsection{RENO}

The RENO $3\nu$ oscillation analysis is based on~\cite{Bak:2018ydk}.
The number of events in the near and far detectors are computed as in
eq.~\eqref{eq:number-of-events}, using the Daya Bay response
function. The average fission fractions are taken
from~\cite{RENO:2018pwo}, the baselines from~\cite{Ahn:2010vy}, and
life time days can be found in~\cite{Bak:2018ydk}.  In order to
compute the total relative efficiency between the near and far
detectors, a normalization to the total number of predicted events
without oscillations in the far detector is performed.  The RENO
$\chi^2$ is based on the far/near spectral ratio is implemented as
follows:
\begin{multline}
  \label{eq:RENO-chisq}
  \chi^2(\theta_{12},\theta_{13},\Dmq_{21},\Dmq_{31}, \vec\eta) \\
  = \sum_i \frac{\left( \frac{O^F_i}{O^N_i} - \frac{N^F_i}{N^N_i}(\theta_{12},\theta_{13},\Dmq_{21},\Dmq_{31}, \vec\eta)\right)^2}{\left( \sigma^\text{stat}_i\right)^2}
  + \vec\eta^T V_\eta^{-1} \vec\eta \,.
\end{multline}
Here, $O^{F/N}_i$, are the observed number of events in the far and
near detectors and energy bin $i$, with background expectations
subtracted. The data is obtained from digitizing fig.~1
of~\cite{Bak:2018ydk}.  $N^{F/N}_i$, are the corresponding predicted
number of events, computed as in eq.~\eqref{eq:number-of-events} and
$\sigma^\text{stat}_i$, are the statistical errors of the ratio
$\frac{O^{F}_i}{O^{N}_i}$. $\vec\eta$ is the vector of the pull
parameters and $V_\eta$ is the pull correlation matrix which accounts
for the systematic uncertainties associated to each pull parameter and
their correlations.

Ref.~\cite{Bak:2018ydk} quotes systematic uncertainties in the
relative detection efficiency and relative energy scale of 0.21\% and
0.15\%, respectively.  The cosmogenic Li-He background plays no
significant role, but is included as well with a relative uncertainty
of 5\% (8\%) for the near (far) detector, cf.\ table~I
of~\cite{Bak:2018ydk}.  In order to match precisely the results shown
in fig.~3 of~\cite{Bak:2018ydk}, an extra factor of 0.984 to the
Far/Near ratio has to be included and the relative detection
efficiency uncertainty has to be increased by a factor of 1.4. Our
re-analysis of RENO data is compared to the official one in the right
panel of fig.~\ref{fig:reactor-check}.

\bibliographystyle{JHEP}
\bibliography{references}

\end{document}